\documentstyle[11pt]{article}
\catcode`@=11
\def\seceqaa{\@addtoreset{equation}{section}
\def\theequation{A\arabic{equation}}}
\def\seceqbb{\@addtoreset{equation}{section}
\def\theequation{B\arabic{equation}}}
\catcode`@=12
\begin{document}
\begin{center}
{\Large \bf
MQCD, (`Barely') $G_2$ Manifolds and (Orientifold of) a Compact Calabi-Yau}
\footnote{This article is partly based on talks given at ``Seventh Workshop on
QCD" [session on ``Strings, Branes and (De-)Construction"], 
Jan 6-10, 2003, La Cittadelle, Villefranche-sur-Mer, France; Fourth
Workshop on ``Gauge Fields and Strings", Feb 25-Mar 1, 2003, Jena, Germany;
``XII Oporto Meeting on Geometry, Topology and Strings", July 17-20, 2003,
Oporto, Portugal; ``SQS03" - International Workshop on ``Supersymmetries and
Quantum Symmetries', July 24-29, 2003, JINR, Dubna, Russia; poster presented
at ``XIV International Congress on Mathematical Physics", July 28-Aug 2, 2003,
Lisbon, Portugal}
\vskip 0.1in
{Aalok Misra\\
Indian Institute of Technology Roorkee,\\
Roorkee - 247 667, Uttaranchal, India\\email: aalokfph@iitr.ernet.in}
\vskip 0.5 true in
\end{center}

\begin{abstract}
We begin with a discussion on two apparently disconnected topics - one 
related to nonperturbative superpotential generated from wrapping an $M2$-brane 
around a supersymmetric three cycle embedded in a $G_2$-manifold 
evaluated by the path-integral inside a path-integral approach of 
\cite{BBS}, and the other centered around the compact Calabi-Yau $CY_3(3,243)$ 
expressed as a blow-up
of a degree-24 Fermat hypersurface in ${\bf WCP}^4[1,1,2,8,12]$. For the former,
we compare the results with the ones of Witten on heterotic world-sheet
instantons \cite{Witten1}. The subtopics covered in the latter include
an ${\cal N}=1$ triality between Heterotic, $M$- and $F$-theories, 
evaluation of ${\bf RP}^2$-instanton superpotential, Picard-Fuchs equation
for the mirror Landau-Ginsburg model corresponding to $CY_3(3,243)$,
$D=11$ supergravity corresponding to $M$-theory compactified on a `barely'
$G_2$ manifold involving $CY_3(3,243)$ and a conjecture related to the 
action of antiholomorphic involution on period integrals.
We then shown an indirect connection between the two topics by showing
a connection between each one of the two and Witten's MQCD\cite{MQCD1}.
As an aside, we show that in the limit of vanishing ``$\zeta$",
a complex constant that appears in the Riemann surfaces relevant to
definining the boundary conditions for the domain wall in MQCD, 
the infinite series of \cite{Vol2} used to represent a suitable embedding
of a supersymmetric 3-cycle in a $G_2$-mannifold,  can be summed.
\end{abstract}

\section{Introduction}

Nonperturbative aspects of string theory have continued to be an extremely
active area of work that bring about a very interesting interplay of various
topics in field theory and algebraic geometry. We will be concentrating
on two such topics - membrane instanton superpotentials in $M$ theory
compactified on $G_2$-manifolds\cite{AM}, and aspects of string and $M$-theory 
compactifications on manifolds involving the compact Calabi-Yau $CY_3(3,243)$
\cite{AM1,AM2,AM3}.
We then attempt to establish an indirect connection between these two
by showing a connection between both and Witten's MQCD\cite{MQCD1}, 
individually. 

In 
Section {\bf 1}, we discuss evaluation of membrane instanton superpotential
and the comparison with Witten's heterotic world-sheet instanton superpotential
\cite{Witten1}. In Section {\bf 2}, we begin with a discussion on 
an ${\cal N}=1,D=4$  Heterotic/$M$/$F$ triality, followed by the Picard-Fuchs
equation derived and solved for the mirror Landau-Ginsburg model for
type $IIA$ compactified on $CY_3(3,243)$, as well as a discussion on 
unoriented world-sheet instanton superpotential, $D=11$ supergravity
corresponding to $M$ theory compactified on the 'barely' $G_2$-manifold
${CY_3(3,243)\times S^1\over{\bf Z}_2}$ and finally a conjecture related
to a freely acting antiholomorphic involution on the period integrals 
for $CY_3(3,243)$. In Section {\bf 3}, we discuss the connection between
sections {\bf 2} and {\bf 3} and Witten's MQCD, individually. Section {\bf 4}
has the conclusions.

\section{Evaluation of the membrane instanton contribution to the superpotential}

String and $M$ theories on manifolds with $G_2$ and $Spin(7)$ holonomies 
have become an active area of research, after construction  of explicit
examples of such manifolds by Joyce\cite{J}. Some explicit metrics
of noncompact manifolds with the above-mentioned exceptional holonomy groups
have been constructed \cite{C}.

Gopakumar and Vafa in \cite{GopVafa}, had conjectured that similar to the large $N$
Chern-Simons/open topological string theory duality of Witten, large
$N$ Chern-Simons on $S^3$ is dual to closed type-A topological string
theory on an $S^2$-resolved conifold geometry. This conjecture was verified
for arbitrary genus $g$ and arbitrary t'Hooft coupling.
This duality was embedded by Vafa in type IIA, and 
was uplifted to $M$ theory on a $G_2$
holonomy manifold by Atiyah, Maldacena and Vafa\cite{amv}.  
The $G_2$ holonomy manifold that was considered by Atiyah et al in \cite{amv} 
was a spin bundle over $S^3$ with the topology of $R^4\times S^3$. 

It will be interesting to be able to lift 
the above-mentioned Gopakumar-Vafa duality to $M$ theory on a $G_2$-holonomy
manifold, and in the process possibly get a formulation of a topological
$M$-theory. As the type-A topological string  theory's partition function
receives contributions only from holomorphic maps from the world-sheet
to the target space, and apart from constant maps, instantons fit the bill,
as a first step we should look at obtaining the superpotential contribution
of multiple wrappings of $M2$ branes on supersymmetric 3-cycles in
a suitable $G_2$-holonomy manifold(membrane instantons). A sketch of the
result anticipated for a single $M2$ brane wrapping an isolated
supersymmetric 3-cycle, was given by Harvey-Moore. In this work, we have worked
out the exact expression for the same, using techniques developed in
\cite{Ovrutetal} on evaluation of  the nonperturbative contribution to the
superpotential of open membrane instantons obtained by wrapping the $M2$
brane on an interval [0,1] times (thus converting the problem 
to that of a heterotic string wrapping) 
a holomorphic curve in a Calabi-Yau three-fold.

As given in \cite{harvmoore}, the Euclidean action for an $M2$ brane is
given
by the following Bergshoeff, Sezgin, Townsend action:
\begin{equation}
\label{eq:EucM2}
{\cal S}_\Sigma=\int d^3z\Biggl[{\sqrt{g}\over l_{11}^3}
-{i\over 3!}\epsilon^{ijk}
\partial_i{\bf Z}^M\partial_j{\bf Z}^N\partial_k{\bf Z}^PC_{MNP}
(X(s),\Theta(s))\Biggr],
\end{equation}
where ${\bf Z}$ is the map of the $M2$ brane world-volume to the
the $D=11$ target space $M_{11}$, both being regarded as supermanifolds.
The $g$ in (\ref{eq:EucM2}), is defined as:
\begin{equation}
\label{eq:gdef}
g_{ij}=\partial_i{\bf Z}^M\partial_j{\bf Z}^N{\bf E}^A_M{\bf
E}^B_N\eta_{AB},
\end{equation}
where ${\bf E}^A_M$ is the supervielbein, given in \cite{harvmoore}.
$X(s)$ and $\Theta(s)$ are the bosonic and fermionic coordinates
of ${\bf Z}$. After using the static gauge and 
$\kappa$-symmetry fixing, the physical degrees of freedom, are given
by $y^{m^{\prime\prime}}$, the section of the normal bundle to
the $M2$-brane world volume, and $\Theta(s)$, section of the 
spinor bundle tensor product: $S(T\Sigma)\otimes S^-(N)$, where
the $-$ is the negative $Spin(8)$ chirality, as under an orthogonal 
decomposition of $TM_{11}|_\Sigma$ in terms of tangent and normal 
bundles, the structure group $Spin(11)$ decomposes into
$Spin(3)\times Spin(8)$.

The action in (\ref{eq:EucM2}) needs to be expanded up to $O(\Theta^2)$,
and the expression is (one has to be careful that in Euclidean $D=11$, one does
not have a Majorana-Weyl spinor or a Majorana spinor) given as:
\begin{eqnarray}
\label{eq:actexpTh2}
& & {\cal S}_\Sigma=\int_\Sigma\Biggl[C +{i\over l_{11}^3}vol(h)
+{\sqrt{h_{ij}}\over l_{11}^3}\biggl(h^{ij}
D_iy^{m^{\prime\prime}} D_j y^{n^{\prime\prime}}
h_{m^{\prime\prime}n^{\prime\prime}}-y^{m^{\prime\prime}}
{\cal U}_{m^{\prime\prime}n^{\prime\prime}}y^{n^{\prime\prime}}+O(y^3)
\biggr)\nonumber\\
& & +{i\over l_{11}^3}\sqrt{h_{ij}}{1\over2}({\bar\Psi}_MV^M 
- {\bar V}^M\Psi_M)+2{\sqrt{h_{ij}}\over l_{11}^3}h^{ij}
{\bar\Theta}\Gamma_iD_j\Theta + O(\Theta^3)\Biggr],
\end{eqnarray}
where we follow the conventions of \cite{harvmoore}: $V_M$ being
the gravitino vertex operator, $\Psi$ being the gravitino field that
enters via the supervielbein ${\bf E}^A_M$, ${\cal U}$ is a mass
matrix defined in terms of the Riemann curvature tensor and the
second fundamental form (See (\ref{eq:Udieuf})).

After $\kappa$-symmetry fixing, like \cite{harvmoore}, we set
$\Theta_2^{A\stackrel{.}{a}}(s)$, i.e., the positive $Spin(8)$-chirality to zero,
and following \cite{Ovrutetal}, will refer to $\Theta_1^{Aa}(s)$ as
$\theta$.

The Kaluza-Klein reduction of the $D=11$ gravitino is given by:
$dx^M\Psi_M=dx^\mu\Psi_\mu+dx^m\Psi_\mu,
\Psi_\mu(x,y)=\psi_\mu(x)\otimes\vartheta(y)$,
$\Psi_m(x,y)=l_{11}^3\sum_{I=1}^{b_3}
\omega^{(3)}_{I,mnp}(y)\Gamma^{pq}\chi^I(x)\otimes\tilde{\eta}(y)$,
where we do not write the terms obtained by expanding in terms of
$\{\omega^{(2)}_{I,mn}\}$, the harmonic 2-forms forming a basis for
$H^2(X_{G_2},{\bf Z})$, as we will be interested in $M2$ branes
wrapping supersymmetric 3-cycles in the $G_2$-holonomy manifold.
For evaluating the nonperturbative contribution to the superpotential,
following \cite{harvmoore}, we will evaluate the fermionic 2-point
function: $\langle\chi^i(x_1^u)\chi^j(x_2^u)\rangle$ (where $x_{1,2}$ are the
${\bf R}^4$ coordinates and u [and later also $v$]$\equiv 7,8,9,10$
is [are] used to index these coordinates), 
and drop the interaction terms in the $D=4,{\cal N}=1$ supergravity action. 
The corresponding mass
term in the supergravity action appears as $\partial_i\partial_j W$,
where the derivatives are evaluated w.r.t. the complex scalar
obtained by the Kaluza-Klein reduction of $C+{i\over l_{11}^3}\Phi$ using 
harmonic three forms forming a basis for $H^3(X_{G_2},{\bf R})$.
One then integrates twice to get the expression for the superpotential
from the 2-point function.

The bosonic zero modes are the four bosonic coordinates that specify
the position of the supersymmetric 3-cycle, and will be denoted
by $x_0^{7,8,9,10}\equiv x^u_0$. The fermionic zero modes come from
the fact that for every $\theta_0$ that is the solution to the fermionic
equation of motion, one can always shift $\theta_0$ to
$\theta_0+\theta^\prime$
, where $D_i\theta^\prime=0$. This $\theta^\prime=\vartheta\otimes\eta$
where $\vartheta$ is a $D=4$ Weyl spinor, and $\eta$ is a covariantly
constant spinor on the $G_2$-holonomy manifold.

After expanding the $M2$-brane action in fluctuations about  solutions to
the bosonic and fermionic equations of motion, one gets that:
${\cal S}|_\Sigma={\cal S}^y_0+{\cal S}^\theta_0+{\cal S}^y_2
+{\cal S}^\theta_2$,
where 
${\cal S}^y_0\equiv {\cal S}_\Sigma|_{y_0,\theta_0}$
${\cal S}^\theta_0\equiv {\cal S}_\Sigma^\theta
+{\cal S}^{\theta^2}_\Sigma|_{y_0,\theta_0}$;
${\cal S}^y_2
\equiv {\delta^2{\cal S}_\Sigma\over\delta y^2}|_{y_0,\theta_0=0}
(\delta y)^2$;
${\cal S}^\theta_2
\equiv {\delta^2{\cal S}\over\delta\theta^2}|_{y_0,\theta_0=0}
(\delta\theta)^2$.
Following \cite{Ovrutetal}, we consider
classical values of coefficients of $(\delta y)^2,(\delta\theta)^2$ terms,
as fluctuations are considered to be of ${\cal O}(\sqrt{\alpha^\prime})$.

Now, 
\begin{eqnarray}
\label{eq:2ptdef}
& & \langle \chi^i(x_1^u)\chi^j(x_2^u)\rangle=\nonumber\\
& & \int {\cal D}\chi e^{K.E\ of\ \chi}\chi^i(x)\chi^j(x)
\int d^4x_0 e^{-{\cal S}_0^y}\nonumber\\
& & \times\int d\vartheta^1d\vartheta^2 e^{-{\cal S}^\theta_0}
\int{\cal D}\delta y^{m^{\prime\prime}}e^{-{\cal S}^y_2}
\int{\cal D}\delta{\bar\theta}{\cal D}\delta\theta
e^{-{\cal S}^\theta_2}.
\end{eqnarray}

We now evaluate the various integrals that appear in (\ref{eq:2ptdef}) above
starting with $\int d^4x e^{-{\cal S}_0^y}$:
\begin{equation}
\label{eq:bzmint}
\int d^4x_0 e^{-{\cal S}_0^y}=\int d^4x_0 e^{[iC-{1\over l_{11}^3}vol(h)]}.
\end{equation}

Using the 11-dimensional Euclidean representation of the gamma matrices
as given in \cite{harvmoore},
${\cal S}^\theta_0+{\cal S}^{\theta^2}_0|_\Sigma={i\over
2l_{11}^3}\int_\Sigma
\sqrt{h_{ij}}{\bar\Psi}_MV_M d^3s,$
where using $\partial_i x_0^u=0$, and using $U$ to denote coordinates
on the $G_2$-holonomy manifold, $V^U=h_{ij}\partial_i y_0^U\partial_j y^V
\gamma_V\theta_0+{i\over2}\epsilon^{ijk}\partial_iy_0^U
\partial_jy_0^V\partial_k y_0^W\Gamma_{VW}\theta_0$,
\begin{equation}
\label{eq:linth}
\int d\vartheta_1 d\vartheta_2e^{{i\over 2l_{11}^3}\sum_{I=1}^{b_3}
\sum_{\alpha=1}^2\sum_{i=1}^8
({\bar\chi}(x)\sigma^{(i)})_\alpha\vartheta_\alpha\omega_I^{(i)}}
=-{1\over 4l_{11}^3}\sum_{I=1}^{b_3}\sum_{i<j=1}^8\omega_I^{(i)}
\omega_I^{(j)}({\bar\chi}\sigma^{(i)})_1({\bar\chi}\sigma^{(j)})_2,
\end{equation}
where one uses that for $G_2$-spinors, the only non-zero 
bilinears are: $\eta^\dagger\Gamma_{i_1...i_p}\eta$ for
$p=0(equiv$ constant), $p=3(\equiv$ calibration 3-form), $p=4(\equiv$)
Hodge dual of the calibration 3-form and $p=7(\equiv$ volume form).
We follow the following notations for coordinates:
$u,v$ are ${\bf R}^4$ coordinates, $\hat{U},\hat{V}$ 
are $G_2$-holonomy manifold
coordinates that are orthogonal to the $M2$ world volume (that wraps
a supersymmetric 3-cycle embedded in the $G_2$-holonomy manifold),
and $U,V$ are $G_2$-holonomy manifold coordinates. The 
tangent/curved space coordinates for $\Sigma$
are represented by $a^\prime/m^\prime$ and those for
$X_{G_2}\times{\bf R}^4$ are represented by 
$a^{\prime\prime}/m^{\prime\prime}$.

We now come to the evaluation of ${\cal S}^\theta_2|_{y_0,\theta_0=0}$.
Using the equality of the two $O((\delta)\Theta^2)$ terms 
in the action of Harvey
and Moore, and arguments similar to the ones in \cite{Ovrutetal}, one can show
that one needs to evaluate the following bilinears:
$\delta{\bar\Theta}\Gamma_{a^\prime}\partial_i\delta\Theta$,
$\delta{\bar\Theta}\Gamma_{a^{\prime\prime}}\partial_i\delta\Theta$,
$\delta{\bar\Theta}\Gamma_{a^\prime}\Gamma_{AB}\delta\Theta$, and
$\delta{\bar\Theta}\Gamma_{m^{\prime\prime}}\Gamma_{AB}\delta\Theta$.
Evaluating them, one gets:
\begin{eqnarray}
\label{eq:th2}
& & {\cal S}^\theta_2|_{y_0,\theta_0=0}\equiv\int_\Sigma d^3s \delta
\theta^\dagger{\cal O}_3\delta\theta,\nonumber\\
& & {\rm where}\ {\cal O}_3\equiv\nonumber\\
& & {2i\over l_{11}^3}\sqrt{h_{ij}}\Biggl[h^{ij}\delta^{m^\prime}_j
\biggl(-e^3_{m^\prime}\sigma^3\otimes{\bf 1}_8\partial_i
-2i[e^1_{m^\prime}\omega^{23}_i+e^2_{m^\prime}\omega^{31}_i+e^3_i\omega^{1
2}_i]
\sigma^3\otimes{\bf 1}_8\nonumber\\
& & -2[e^1_{m^\prime}\omega^{13}_i+e^2_{m^\prime}\omega^{23}_i]
\sigma^3\otimes{\bf 1}_8
+e^3_{m^\prime}{\omega^{b^{\prime\prime}c^{\prime\prime}}\over2}\sigma^3\otimes
\gamma_{b^{\prime}c^{\prime}}\biggr)\nonumber\\
& & +ih^{ij}e^{a^\prime}_{m^\prime}\omega^{b^\prime c^{\prime\prime}}_i
\sigma^2\otimes\gamma_{c^{\prime\prime}}[\delta^{a^\prime b^\prime}
+i\epsilon^{a^\prime b^\prime c^\prime}\delta^{c^\prime}_3]\nonumber\\
& & +ih^{ij}\partial_j y^{m^{\prime\prime}}
e^{a^{\prime\prime}}_{m^{\prime\prime}}[
\sigma^2\otimes\gamma_{a^{\prime\prime}}\partial_i
-(\omega^{b^{\prime\prime}c^{\prime
\prime}}_i\sigma^2{1\over6}
\otimes\gamma_{a^{\prime\prime}b^{\prime\prime}c^{\prime\prime}}
-{1\over2}\omega^{b^\prime c^{\prime\prime}}\delta^{b^\prime}_3
\sigma^3\otimes\gamma_{a^{\prime\prime}c^{\prime\prime}})]\Biggr],\nonumber\\
& & 
\end{eqnarray}
Hence, the integral over the fluctuations in $\theta$ will give a
factor of $\sqrt{det{\cal O}_3}$ in Euclidean space. 

The expression for ${{\cal S}^y_2|_\Sigma}_{y_0,\theta_0=0}$ is identical
to the one given in \cite{Ovrutetal}, and will contribute 
${1\over{\sqrt{det{\cal O}_1det{\cal O}_2}}}$, where ${\cal O}_1$ and
${\cal O}_2$ are as given in the same paper:
\begin{eqnarray}
\label{eq:bosondets}
& & {\cal O}_1\equiv\eta_{uv}\sqrt{g}g^{ij}{\cal D}_i\partial_j\nonumber\\
& & {\cal O}_2\equiv\sqrt{g}(g^{ij}{\cal D}_ih_{\hat{U}\hat{V}}
D_j+{\cal U}_{\hat{U}\hat{V}}).
\end{eqnarray}
The mass matrix ${\cal U}$ is expressed in terms of the curvature
tensor and product of two second fundamental forms.
${\cal D}_i$ is a covariant derivative with indices in the corresponding
spin-connection of the type $(\omega_i)^{m^{\prime\prime}}_{n^{\prime\prime}}$
and $(\omega_i)^{m^\prime}_{n^\prime}$, and $D_i$ is a covariant derivative
with  corresponding spin connection indices of the former type.

Hence, modulo supergravity determinants, and the contribution from the
fermionic zero modes, the exact form of the superpotential
contribution coming from a single $M2$ brane wrapping an isolated supersymmetric
cycle of $G_2$-holonomy manifold, is given by:
\begin{equation}
\label{eq:Wfinal}
\Delta W = e^{iC - {1\over l_{11}^3}vol(h)}\sqrt{{det O_3\over
det {\cal O}_1\ det{\cal O}_2}}.
\end{equation}
We do not bother about 5-brane instantons, as
we assume that there are no supersymmetric 6-cycles in the $G_2$-holonomy
manifold that we consider.

In \cite{harvmoore}, it is argued that for 
an ``associative'' 3-fold $\Sigma$
in the $G_2$-holonomy manifold, the structure group $Spin(8)$ decomposes
into $Spin(4)_{\bf R^4}\times Spin(4)_{X_{G_2}\setminus\Sigma}$. After
gauge-fixing under $\kappa$-symmetry, 
\begin{equation}
\label{eq:thcomps}
\Theta=\biggl((\Theta_{--})^{AY}_\alpha, 
(\Theta_{++})^{\stackrel{\cdot}{Y}}_{\stackrel{\cdot}{\alpha} A};
0,0\biggr),
\end{equation}
where $A,\stackrel{(\cdot)}{\alpha},\stackrel{(\cdot)}{Y}$ 
are the $Spin(3), Spin(4)_{\bf R^4},
 Spin(4)_{X_{G_2}\setminus\Sigma}$ indices respectively. The $G_2$ structure allows
one to trade off $(\Theta_{--})^{AY}_\alpha$  for fermionic 0- and 1-forms:
$\eta,\chi_i$, which together with $y^u\equiv 
y^{\alpha\stackrel{\cdot}{\alpha}}$, form
the Rozansky-Witten(RW) multiplet. Similarly, $(\Theta_{++})^{
\stackrel{\cdot}{Y}}_{\stackrel{\cdot}{\alpha} A}$
gives the Mclean multiplet: $
(y^{A\stackrel{\cdot}{y}},\nu^{\stackrel{\cdot}{Y}}_{\stackrel{\cdot}{\alpha}
 A})$.
The RW model is a $D=3$ topological sigma model on a manifold
embedded in a hyper-K\"{a}hler manifold $X_{4n}$ \cite{RozWitt}. If
$\phi^{M(=1,...,4n)}(x^i)$
are functions from mapping $M$ to $X$, then the RW action is given by:
\begin{eqnarray}
\label{eq:RW}
\int_\Sigma\sqrt{h_{ij}}\Biggl[{1\over2}h_{MN}\partial_i\phi^M\partial_j\phi^N
h^{ij}+\epsilon_{IJ}h^{ij}\chi^I_iD_j\eta^J
+{1\over2\sqrt{h_{ij}}}\epsilon^{ijk}\biggl(\epsilon_{IJ}
\chi^I_iD_j\chi^J_k+{1\over3}\Omega_{IJKL}\chi_i^I\chi_j^J\chi^L_k\eta^L\biggr) 
\Biggr],
\end{eqnarray}
where $\Omega_{IJKL}=\Omega_{JIKL}=\Omega_{IJLK}$. Then, dropping
the term proportional to $\Omega_{IJKL}$, one sees that the terms in 
(\ref{eq:actexpTh2}),  are very likely to give the RW action in
(\ref{eq:RW}).
In \cite{harvmoore}, $n=1$. 

As the RW and Mclean's multiplets are both contained in 
$\delta\theta$, hence (using the notations of \cite{harvmoore})
$det^\prime(L_-)det^\prime(\rlap/D_E)$ will be
given by $det{\cal O}_3$ - it is
however difficult to disentangle the two contributions.
The relationship
involving the spin connections on the tangent bundle and normal bundle (the anti
self-dual part of the latter) as given in \cite{harvmoore}, can be used to
reduce the number of independent components of the spin connection
and thus simplify (\ref{eq:th2}).  
Further, $(det^\prime\Delta_0)^2|det^\prime(\rlap/D_E)|$ 
should be related to $\sqrt{det{\cal O}_1det{\cal O}_2}$. 
Hence, 
the order of $H_1(\Sigma,{\bf Z})$ must
be expressible in terms of $\sqrt{det {\cal O}_{1,2,3}}$ for
$M2-$brane wrapping a rigid supersymmetric 3-cycle. However, we wish to
emphasize that (\ref{eq:th2}), unlike the corresponding result
of \cite{harvmoore}, 
is equally valid for $M2-$brane wrapping a non-rigid supersymmetric 3-cycle,
as considered in Section 4.

We now explore the possibility of cancellations 
between the bosonic and fermionic determinants
For bosonic determinants $det A_b$, the function that is relevant
is $\zeta(s|A_b)$, and that for fermionic determinants $det A_f$, 
the function that is additionally
relevant is $\eta(s|A_f)$. The integral representation of the former involves
$Tr(e^{-tA_b})$, while that for the latter involves $Tr(Ae^{-tA^2})$
(See \cite{elizalde}):
\begin{eqnarray}
\label{eq:zetaetaMellin}
& & \zeta(s|A_b)={1\over\Gamma(2s)}\int_0^\infty dt t^{s-1}
Tr(e^{-tA_b});
\eta(s|A_f)={1\over\Gamma({s+1\over2})}\int_0^\infty
dt t^{{s+1\over2}}Tr(A_fe^{-tA_F^2}) 
, 
\end{eqnarray}
where to get
the UV-divergent 
contributions, one looks at the $t\rightarrow0$ limit of the two terms.        
To be  more precise (See \cite{Deseretal})
\begin{eqnarray}
\label{eq:bosfermdets}
& & ln det A_b = -{d\over ds}\zeta(s|A_b)|_{s=0}\nonumber\\
& & =-{d\over ds}\biggl({1\over\Gamma(s)}\int_0^\infty dt t^{s-1}
Tr(e^{-tA_b})\biggr)|_{s=0};\nonumber\\
& & ln det A_f = -{1\over2}{d\over ds}\zeta(s|A_f^2)|_{s=0}
\mp{i\pi\over 2}\eta(s|A_f)|_{s=0} \pm{i\pi\over 2}\zeta(s|A_f^2)|_{s=0}
\nonumber\\
& & = \biggl[-{1\over2}{d\over ds}\pm{i\pi\over2}\biggr]\biggl(
{1\over\Gamma(s)}\int_0^\infty dt t^{s-1}Tr(e^{-tA_f^2})\biggr)|_{s=0}
\mp{i\pi\over2}{1\over\Gamma({s+1\over2})}\int_0^\infty dt t^{{s+1\over2}-1}
Tr(A_f e^{-tA_f^2})|_{s=0},\nonumber\\
& & 
\end{eqnarray}
where the $\mp$ sign in front of $\eta(0)$, a non-local object, represents
an ambiguity in the definition of the determinant. The $\zeta(0|A_f^2)$ term
can be reabsorbed into the contribution of $\zeta^\prime(0|A_f^2)$, and
hence will be dropped below.  
Here $Tr\equiv\int dx\langle x|...|x\rangle\equiv\int dx tr(...)$. The idea
is that if one gets a match in the Seeley - de Witt coefficients for
the bosonic and fermionic determinants, implying equality of 
UV-divergence, this is indicative of a possible complete cancellation. 

The heat kernel expansions for the bosonic and fermionic determinants\cite{Gilkey}
are given by:
\begin{eqnarray}
\label{eq:heatkernexps}
& & tr(e^{-tA_b})=\sum_{n=0}^\infty e_n(x,A_b) t^{{(n-m)\over2}},
tr(A_fe^{-tA_f^2})=\sum_{n=0}^\infty a_n(x,A_f) t^{{(n-m-1)\over2}},
\end{eqnarray}
where for $m$ is the dimensionality of the space-time. For our case, we have a compact
3-manifold, for which $e_{2p+1}=0$ and $a_{2p}=0$. For Laplace-type operators $A_b$,
and Dirac-type operators $A_f$, the non-zero coefficients are determined to be the
following:
\begin{eqnarray}
\label{eq:e_ns}
& & e_0(x,A_b)=(4\pi)^{-{3\over2}}Id,\
e_2(x,A_b)=(4\pi)^{-{3\over2}}\biggl[\alpha_1 E+\alpha_2\tau Id\biggr],
\end{eqnarray}
where $\alpha_i$'s are constants, $\tau\equiv R_{ijji}$, and
$Id$ is the identity that figures with the scalar leading symbol in the
Laplace-type operator $A_b$ (See \cite{Gilkey}), and
\begin{eqnarray}
\label{eq:Edieuf}
& & 
E\equiv B - G^{ij}(\partial_i\omega_j+\omega_i\omega_j-\omega_k\Gamma^k_{ij}),
\nonumber\\
& & A_b\equiv-(G^{ij}Id\partial_i\partial_j+A^i\partial_i+B),\nonumber\\
& & \omega_i={G_{ij}(a^j+G^{kl}\Gamma^j_{kl}Id)\over 2}.
\end{eqnarray}
To actually evaluate $e_0$ and $e_2$, we need to find an example
of a regular $G_2$-holonomy manifold that is metrically $\Sigma\times M_4$,
where $\Sigma$ is a supersymmetric 3-cycle on which we wrap an $M2$
brane once, and $M_4$ is a four manifold.
 One such example was obtained in \cite{gyz}, 
that be regarded as a cone over a base $S^3\times R^3$,
that I was referring to is actually:
\begin{equation}
\label{eq:GYZ1}
ds^2=dr^2+{1\over y}\sum_{j=1}^3(x^2\sigma_j^2+mn\sigma_j d\theta_j-
mx d\theta^2_j),
\end{equation}
where $\alpha_{1,2,3}$ are the left-invariant $SU(2)$ 1-forms, $m,n$ are
two parameters that characterize $H^3(S^3\times R^3,{\bf R})={\bf
R}\oplus
{\bf R}$, $x=-{m^{{1\over3}}\over4}(r-r_0)^2,
\ y={m^{{2\over3}}\over4}(r-r_0)^2$, $r_0$ being an integration constant
that for convenience can be set to zero.

{\bf Now take the simplifying limit $n=0$}. This for $m=1$ gives:
\begin{equation}
\label{eq:G2metric}
ds_7^2= dr^2+{r^2\over4}\sum_{i=1}^3\sigma_i^2 + \sum_{i=1}^3
d\theta_i^2,
\end{equation}
where $\sigma_i$'s are left-invariant one-forms obeying the SU(2)
algebra: $d\sigma_i=-{1\over2}\epsilon^{ijk}d\sigma^j\wedge d\sigma^k$,
given by:
\begin{eqnarray}
\label{eq:sigmas}
& & \sigma_1\equiv cos\psi d\theta + sin\psi sin\theta d\phi,
\sigma_2\equiv -sin\psi d\theta + cos\psi sin\theta d\phi,
\sigma_3\equiv d\psi + cos\theta d\phi.
\end{eqnarray}
The metric in ( ref{eq:sigmas})
does not have a $G_2$ holonomy. It is argued that (\ref{eq:G2metric}) 
is what (\ref{eq:GYZ1}) asymptotes to, for $n\neq0$.
The heat-kernel asymptotics analysis below,
can either be treated
as one for membrane instanton superpotentials for non-compact $G_2$
manifolds
IN THE LARGE DISTANCE-LIMIT($\leftrightarrow r\rightarrow\infty$), or
equivalently for a non-compact $M_7$ with a supersymmetric 3-cycle(a
$T^3$)
embedded in it that nevertheless gives ${\cal N}=1, D=4$ supersymmetry.

 To see
 that the $T^3$ corresponds to a supersymmetric 3-cycle, we need to 
show that the pull-back of the calibration $\Phi_3$ restricted to $\Sigma$,
is the volume form on $\Sigma$ (See \cite{beckersetal}).
$\Phi_3$ using the notations of \cite{gyz} is given by:
\begin{equation}
\label{eq:caldieuf}
\Phi_3=e^{125} + e^{147} +e^{156}-e^{246} + e^{237}+e^{345} +e^{567},
\end{equation}
where $e^{ijk}\equiv e^i\wedge e^j \wedge e^k$.
Let $1,...,7$ denote $r,\psi,\theta,\phi,\theta_1,\theta_2,\theta_3$.
Hence, when restricted to $\Sigma(\theta_1,\theta_2,\theta_3)$ using
the static gauge, one gets:
$\Phi_3|_\Sigma=e^{567}=d\theta_1\wedge d\theta_2\wedge d\theta_3,$
which is the volume form on $T^3$. Thus, the $T^3$ of 
(\ref{eq:G2metric}) is a supersymmetric 3-cycle.

For (\ref{eq:G2metric}), one sees  that $g_{ij}=\delta_{ij}+
\partial_i y_0^{\hat{U}}\partial_j y_0^{\hat{V}}g_{\hat{U}\hat{V}}$,
having used the definition of $g_{ij}$ as a pull-back of the
space-time metric $g_{MN}$, static gauge and that $\partial_i y_0^u=0$.
If one assumes that the coordinates $r,\psi,\theta,\phi$ are
very slowly varying functions of $\theta_1,\theta_2,\theta_3$,
one sees that $g_{ij}\sim\delta_{ij}$. This simplifies the algebra,
though one can work to any desired order 
in $(\partial_i y_0^{\hat{U}})^{p(>0)}$,
and get conclusions similar to the ones obtained below. 

Lets first consider the Seeley de-Wit coefficients for ${\cal O}_1$.
Now, in the above adiabatic approximation, the world volume metric of
the $M2$-brane is flat. Hence, the Christoffel connection $\Gamma^i_{jk}$
for ${\cal O}_1$, vanishes. Now, 
$\omega^{a^\prime b^\prime}_i\sim\delta^{m^\prime}_i
\omega_{m^\prime}^{a^\prime b^\prime}$, where 
\begin{equation}
\label{eq:spcon}
\omega_{m^\prime}^{ab} = e^{[a}_{\ n^\prime}g^{n^\prime l^\prime}
(\partial_{m^\prime} e_{l^\prime}^{\ b^\prime]} - 
\Gamma^{p^\prime}_{l^\prime m^\prime}e_{p^\prime}^{\ b^\prime]}),
\end{equation}
the antisymmetry indicated on the right hand side of 
(\ref{eq:spcon}) being applicable only to the tangent-space indices
$a^\prime,\ b^\prime$, and 
where for (\ref{eq:G2metric}), the following are the non-zero vielbeins:
\begin{eqnarray}
\label{eq:viels}
& & e^1_{\ r}=1; 
e^2_{\ \theta}={r\over2}cos\psi,\ e^2_{\ \phi}={r\over2}sin\psi sin\theta;
\nonumber\\
& & e^3_{\ \theta}=-{r\over2}sin\psi,\ e^3_{\ \phi}=cos\psi sin\theta;
e^4_{\ \psi}={r\over2},\ e^4_{\ \phi}={r\over2}cos\theta;\nonumber\\
& & e^5_{\ \theta_1}=e^6_{\ \theta_2}=e^7_{\ \theta_3}=1.
\end{eqnarray}
Hence, for the $G_2$ metric of (\ref{eq:G2metric}), 
$\partial_{m^\prime}e_{l^\prime}^{\ b^\prime}=0$. Also, 
$\Gamma^{p^\prime}_{l^\prime m^\prime}=0$. Thus, $\omega^{\cal D}_i\sim 0$.

In the adiabatic approximation, $\tau\sim0$. 

Hence, 
\begin{eqnarray}
& & e_0(x,{\cal O}_1)=(4\pi)^{-{3\over2}}; e_2(x,{\cal O}_1)=0.
\end{eqnarray}

We next consider evaluation of $e_{0,2}(x,{\cal O}_2)$. Once again,
the Christoffel connection $\Gamma^i_{jk}$ vanishes. Again, 
$\omega^{D,{\cal D}}\sim 0$. Also, $\partial_i h_{\hat{U}\hat{V}}\sim0$.
Hence, $A^i\sim 0$.

Now,
\begin{equation}
\label{eq:Udieuf}
{\cal U}_{\hat{U}\hat{V}}\equiv{1\over2} R^{m^\prime}_{\hat{U}m^\prime\hat{V}}
+{1\over8}Q^{m^\prime n^\prime}_{\hat{U}}Q_{m^\prime n^\prime\hat{V}},
\end{equation}
where the second fundamental form is defined via:
\begin{equation}
\label{eq:Qdieuf}
\Gamma^{m^{\prime\prime}}_{k^\prime l^\prime}\equiv -{1\over2}
g^{m^{\prime\prime}n^{\prime\prime}}
Q_{k^\prime l^\prime n^{\prime\prime}}.
\end{equation}
Using:
\begin{equation}
\label{eq:Rdieuf}
R^{m^\prime}_{\hat{U}m^\prime\hat{V}}=
\partial_{\hat{V}}\Gamma^{m^\prime}_{\hat{U}m^\prime}
-\partial_{m^\prime}\Gamma^{m^\prime}_{\hat{U}\hat{V}}
+\Gamma^V_{\hat{U}m^\prime}\Gamma^{m^\prime}_{V\hat{V}}
-\Gamma^V_{\hat{U}\hat{V}}\Gamma^{m^\prime}_{Vm^\prime},
\end{equation}
and the fact that the non-zero Christoffel symbols do not involve
$m^\prime$ as one of the (three) indices and that their values are
$m^\prime$-independent, one sees that 
\begin{equation}
\label{eq:nullU}
{\cal U}_{\hat{U}\hat{V}}=0.
\end{equation}
Hence, $B\sim0$. 

For evaluating $\tau\equiv g^{i_1 i_2}g^{j_1 j_2}R_{i_1 j_1 j_2 i_2}$
$=g^{i_1 i_2}g^{j_1 j_2}g_{i_1 l_1}R^{l_1}_{j_1 j_2 i_2}$, one
needs to evaluate $R^{l_1}_{j_1 j_2 i_2}=\partial_{i_2}\Gamma^{l_1}_{j_1 j_2}$
-$\partial_{j_2}\Gamma^{l_1}_{j_1 i_2}$ $+\Gamma^p_{j_1 j_2}\Gamma^l_{p i_2}$
-$\Gamma^p_{j_1 i_2}\Gamma^{l_1}_{p j_2}$. This will be evaluated using
the metric given by $G^{ij}=g^{ij}\sqrt{g}h_{\hat{U}
\hat{V}}$, 
where we will use the adiabatic approximation:
$g_{ij}\sim\delta^{m^\prime}_i\delta^{n^\prime}_jg_{m^\prime n^\prime}$.
Due to the $\hat{U}\hat{V}$ indices, the Ricci scalar $\tau$ is 
actually a matrix in the $X_{G_2}\setminus\Sigma$ space.
In the adiabatic approximation, only the product of the two 
Christoffel symbols in $R^{l_1}_{j_1 j_2 i_2}$ is non-zero, and is
given by the following expression:
\begin{eqnarray}
\label{eq:tauexp}
& & \Gamma^{p^{\prime\prime}}_{j_1 j_2}\Gamma_{p^{\prime\prime}i_2}^{l_1}
-\Gamma^{p^{\prime\prime}}_{j_2 i_1}\Gamma^{l_1}_{p^{\prime\prime} j_2}
\nonumber\\
& & =-{h_{\hat{U}\hat{V}}\over4}\delta_{j_1 j_2}\delta^{l_1}_{i_2}\Biggl[
\biggl(\partial_r\biggl[{1\over h_{\hat{U}\hat{V}}}\biggr]\biggr)^2
+{4\over r^2}\biggl(\partial_\theta\biggl[{1\over h_{\hat{U}\hat{V}}}\biggr]
\biggr)^2\Biggr]\nonumber\\
& & =-{\delta_{j_1 j_2}\delta^{l_1}_{i_2}\over 4}
\left(\begin{array}{cccc}\\
0 & 0 & 0 & 0 \\
0 & {16\over r^4} & 0 & {16\over r^4 cos\theta}
+{4sin^2\theta\over r^2 cos^3\theta} \\
0 & 0 & {16\over r^4} & 0 \\
0 & {16\over r^4 cos\theta}+{4sin^2\theta\over r^2 cos^3\theta} 
& 0 & {16\over r^4} \\
\end{array}\right).
\end{eqnarray} 
Hence, on taking $tr_{X_{G_2}\setminus\Sigma}$, one gets:
\begin{equation}
\tau={72\over r^4}.
\end{equation}
Hence,
\begin{eqnarray}
& & e_0(x,{\cal O}_2)=(4\pi)^{-{3\over2}};\
e_2(x,{\cal O}_2)=(4\pi)^{-{3\over2}}{72\alpha_2\over r^4}.
\end{eqnarray}

We now do a heat-kernel asymptotics analysis of the fermionic determinant
$det{\cal O}_3$.
The fermionic operator ${\cal O}_3$ can be expressed as:
\begin{eqnarray}
\label{eq:O3}
& & {\cal O}_3\equiv \sqrt{h}h^{ij}\Gamma_jD_i=\sqrt{h}h^{ij}\Gamma_j
\biggl(\partial_i+{1\over4}\omega^{a^\prime b^\prime}_i
\Gamma_{a^\prime b^\prime}
+{1\over4}\omega^{a^\prime b^{\prime\prime}}_i\Gamma_{a^\prime b^{\prime\prime}}
\biggr) \equiv G^{ij}\Gamma_j\partial_i-r,
\end{eqnarray}
where
\begin{eqnarray}
\label{eq:Grdieufs}
& & 
G^{ij}\equiv\sqrt{h}h^{ij};
r\equiv{-1\over4}\sqrt{h}h^{ij}\Gamma_j\biggl(
\omega^{a^\prime b^\prime}_i\Gamma_{a^\prime
b^\prime}
+\omega^{a^\prime b^{\prime\prime}}_i\Gamma_{a^\prime b^{\prime\prime}}\biggr).
\end{eqnarray}
${\cal O}_3$ is of the Dirac-type as ${\cal O}_3^2$ is of the Laplace-type,
as can be seen from the following:
\begin{eqnarray}
\label{eq:O3squared}
& & {\cal O}_3^2\equiv G^{ij}\partial_i\partial_j+A^i\partial_i+B,\ {\rm where}:
\nonumber\\
& & G^{ij}\equiv hh^{ij};\nonumber\\
& & A^i\equiv\sqrt{h}\Gamma^j\Gamma_k\partial_j(\sqrt{h}h^{kl})
+2h\Gamma^i\Gamma^l\omega_l^{CD}\Gamma_{CD};\nonumber\\
& & B\equiv\sqrt{h}\Gamma^j\Gamma_k\Gamma_{CD}\partial_j(\sqrt{h}h^{kl}\omega_l^{CD})
+h\Gamma^j\Gamma^l\omega_j^{AB}\Gamma_{AB}\omega_l^{CD}\Gamma_{CD}.
\end{eqnarray}
Now, 
\begin{equation}
\label{eq:phi1}
{\cal O}_3\equiv 
G^{ij}\Gamma_j\bigtriangledown_i -\phi,
\end{equation}
where
$\phi\equiv r+\Gamma^i\omega_i$, and
\begin{eqnarray}
\label{eq:omegacondieuf}
& & \omega_l\equiv{G_{il}\over2}(-\Gamma^j\partial_j\Gamma^i+\{r,\Gamma^i\}+G^{jk}\Gamma^i_{jk})
\nonumber\\
& & ={h_{il}\over2\sqrt{h}}\biggl({1\over4}\sqrt{h}h^{i^\prime j^\prime}
\{\Gamma_{j^\prime}(\omega^{a^\prime b^\prime}_{i^\prime}\Gamma_{a^\prime
b^\prime}
+\omega^{a^\prime b^{\prime\prime}}_{i^\prime}
\Gamma_{a^\prime b^{\prime\prime}}),
\Gamma^i\}\nonumber\\
& & 
+{hh^{jk}h^{ii^\prime}
\over2}\biggl(\partial_j\biggl[{h_{ki^\prime}\over\sqrt{h}}\biggr]
+\partial_k\biggl[{h_{ji^\prime}\over\sqrt{h}}\biggr]
-\partial_{i^\prime}\biggl[
{h_{jk}\over\sqrt{h}}\biggr]\biggr)\biggr).
\end{eqnarray}

The Seeley-de Witt coefficients $a_i$ are given by (See \cite{Gilkey}):
\begin{equation}
\label{eq:fermionais}
a_1(x,G^{ij}\Gamma_j\bigtriangledown_i -\phi)=-(4\pi)^{-{3\over2}}tr(\phi);
a_3(x, G^{ij}\Gamma_j\bigtriangledown_i -\phi)=-{1\over6}(4\pi)^{-{3\over2}}
tr(\phi\tau+6\phi{\cal E}-\Omega_{a^\prime b^\prime;a^\prime}
\Gamma_{b^\prime}),
\end{equation}
where
\begin{equation}
\label{eq:cal Edieuf}
{\cal E}\equiv-{1\over2}\Gamma^i\Gamma^j\Omega_{ij}+\Gamma^i\phi_{;i}-\phi^2,
\Omega_{ij}\equiv\partial_i\omega_j-\partial_j\omega_i+[\omega_i,\omega_j].
\end{equation}
Now, e.g., $\Gamma^i=\partial^i y^M\Gamma_M$, 
where $y^M\equiv y^{m^\prime,\hat{U},u}$
and given that $\partial_i y^u=0$, then in the static gauge,
$\Gamma^i=\delta^i_{m^\prime}\Gamma_{m^\prime}+\partial^i y^{\hat{U}}
\Gamma_{\hat{U}}$
$=\delta^i_{m^\prime}e_{m^\prime}^{\ a^\prime}\Gamma_{a^\prime}$
+ $\partial^i y^{\hat{U}} e_{\hat{U}}^{\ \hat{A}}\Gamma_{\hat{A}}$. Now, 
lets make the simplifying
assumption as done for the bosonic operators, we assume that 
$y^{\hat{U}}$ varies very slowly w.r.t. the $M2$-brane world-volume coordinates.
Hence, we drop all terms of the type $(\partial_i y^{\hat{U}})^{p(>0)}$.
The conclusion below regarding the vanishing of the Seeley-de Witt
coefficients $a_1$ and $a_3$, will still be valid.
The dropping of $(\partial_i y^U)^{p(>0)}$-type terms
will be indicated by $\sim$ as opposed to $=$ in the equations below.
One finally gets: 
\begin{equation}
\label{eq:SdWferm1}
a_1(x, G^{ij}\Gamma_j\bigtriangledown_i -\phi)
=a_3(x, G^{ij}\Gamma_j\bigtriangledown_i -\phi) \sim0.
\end{equation}
We conjecture that in fact, $a_{2n+1}(x,G^{ij}\Gamma_j\bigtriangledown_i-\phi)\sim0$
for $n=0,1,2,3,...$.

By using reasoning similar to the one used in Appendix A, one can show that:
\begin{eqnarray}
\label{eq:O3sqzetaSdW}
& & e_0(x,{\cal O}_3^2)=(4\pi)^{-{3\over2}}; e_2(x,{\cal O}_3^2)\sim0.
\end{eqnarray}
From the extra factor of ${1\over2}$ multiplying the 
$\zeta^\prime(0|{\cal O}_3^2)$ relative to $\zeta^\prime(0|{\cal O}_1)$ in
(\ref{eq:bosfermdets}), and (\ref{eq:SdWferm1}) and (\ref{eq:O3sqzetaSdW}),
one sees the possibility that:
\begin{equation}
\label{eq:bosfermcan}
{ln[det {\cal O}_3]\over ln[det {\cal O}_1]}\sim{1\over2}.
\end{equation}

In conclusion, one sees that Seeley-de Witt coefficients of the
fermionic operator ${\cal O}_3$ are proportional to  those of  
the bosonic operator
 ${\cal O}_1$ in the adiabatic approximation, to the order calculated,
 for the $G_2$-metric 
(\ref{eq:G2metric}). This is indicative of
possible cancellation between them. This is expected, as the $M2$-brane action
has some supersymmetry. As $b_1(T^3)=3>0$, thus the supersymmetric
3-cycle of (\ref{eq:G2metric}) is an example of a non-rigid 
supersymmetric 3-cycle. The result of \cite{harvmoore} is not applicable
for this case. On the other hand, the superpotential written out
as determinants, as in this work, is still valid.
The corresponding modified formula in \cite{harvmoore} might consist, as 
prefactors, in addition to the phase, the torsion elements of
$H_1(\Sigma,{\bf Z})$, represented by $|H_1(\Sigma,{\bf Z})|^\prime$
in \cite{RozWitt}, and perhaps a geometrical quantity that would encode
 the $G_2$-analog of the arithmetic genus condition 
in the context of 5-brane instantons obtained by wrapping $M5$-brane
on supersymmetric 6-cycles in $CY_4$ in \cite{Witten}.
The validity of the arithmetic genus argument, even for $CY_4$,
 needs to be independently verified though.

For heterotic world sheet instantons, as studied in \cite{Witten1}, the
expression for the nonperturbative superpotential is given by:
\begin{equation}
\label{eq:sup1}
\Delta W=exp[-{A(C)\over2\pi\alpha^\prime}+\int_C B]
{{\rm Pfaff}^\prime\partial_{S_+\otimes S_+(N)}
{\rm Pfaff}({\bar\partial}_{O(-1)\otimes V}\over (det\partial_{O(-1)})^2)
({\rm det}\partial_{O(0)})^2({\rm det}{\bar\partial}_{{\cal O}(-1)})^2
({\rm det}^\prime{\bar\partial}_{{\cal O}(0)})^2}
\end{equation}
In this
expression, the convention followed by Witten is that the left fermionic
movers come with the kinetic operator ${\bar\partial}$ and the right
movers 
 come with the kinetic operator ${\partial}$. Further, the bosonic zero
modes are associated with the left-movers corresponding to translation in
${\bf R}^4$, and the fermionic zero modes are associated with the right-movers.
The left-moving fermions are sections: $\Gamma[S_{-}(C\hookrightarrow CY_3)
\otimes V]$
and the right-moving fermions are sections: $\Gamma[S_+(C\hookrightarrow CY_3)
\otimes S_+(N)]$, where $C$ is the genus-zero curve around which the 
string world-sheet wraps around to give world-sheet instanton, $V$ is the
$SO(32)$ gauge bundle, and $N$ is
the normal bundle to $C$ in ${\bf R}^4\times CY_3$. As fiber bundles,
$S_-(C)\cong{\cal O}(-1)$, hence the left movers are sections of
${\cal O}(-1)\otimes V\equiv V(-1)$. Now, $N|_{CY_3}\cong{\cal O}(-1)\oplus
{\cal O}(-1)$, and $N|_{{\bf R}^4}\cong{\cal O}(0)\oplus{\cal O}(0)$. Also,
the eight real bosons transverse to $C$ can be combined into four complex
bosons.

After the bosonic-fermionc determinant cancelation:
\begin{equation}
\label{eq:susy}
{\rm Pfaff}^\prime\partial_{S_+\otimes S_+(N)}
=det\partial_{{\cal O}(-1)})^2 (det\partial_{{\cal O}(0)})^2,
\end{equation}
 what survives in the heterotic world-sheet
instanton superpotential is 
\begin{equation}
\label{eq:sup2}
\Delta W=Exp[-{A(C)\over2\pi\alpha^\prime}+i\int_C B]
{{\rm Pfaff}({\bar\partial}_{V(-1)})\over 
({\rm det}{\bar\partial}_{{\cal O}(-1)})^2
({\rm det}^\prime{\bar\partial}_{{\cal O}(0)})^2}.
\end{equation}
Its here that the difference between the present result and the form
of the result in \cite{Witten1} becomes manifest. In the latter,
after equality of bosonic and fermionic determinants 
one gets Pfaff$({\bar\partial}_{O(-1)\otimes V}$ in the numerator. 
As argued in \cite{Witten1}, this  expression can vanish if the gauge 
bundle restricted to the genus-0 curve $C$, is trivial. To see this,
Witten argued that any $SO(32)$ gauge bundle $V$ over a genus-zero curve $C$,
can be written as:$V=\oplus_{i=1}^{16}{\cal O}(m_i)\oplus {\cal O}(-m_i)$,
implying $V(-1)=\oplus_{i=1}^{16}{\cal O}(m_i-1)\oplus{\cal O}(-m_i-1)$.
Now, dim[ker$({\bar\partial}_{{\cal O}(s)}$]]=$s+1$ if $s>0$ and zero if
$s<0$. Thus, dim[ker$({\bar\partial}_{V(-1)})]=\sum_{i=1}^{16}m_i\equiv0$
iff $m_i=0$ for all $i=1,...,16$.
In our result for the membrane
instanton for non-rigid supersymmetric 3-cycle, the superpotential can never
vanish because of unity in the numerator.

\section{String and $M$-Theory Compactifications involving $CY_3(3,243)$}

In this section, we will discuss two topics. One will be related to an
${\cal N}=1, D=4$ triality between Heterotic, $M$ and $F$ theories. 
The other is related to studying some algebraic geometric aspects of 
the compact Calabi-Yau $CY_3(3,243)$ such as period integrals.

\subsection{An ${\cal N}=1$ Triality by Spectrum Matching}

As ${\cal N}=1$ supersymmetry in four dimensions is of phenomenological interest, it 
is important to understand possible dualities between different ways of arriving at
the same amount of supersymmetry via suitable compactifications. In this 
regard, the results of \cite{VW,AW} are of
particular interest. While \cite{VW} construct such string dual pairs,
\cite{AW} also give ${\cal N}=1$ Heterotic/$M$-theory dual pairs.
As $M$-theory on $G_2$-holonomy manifolds gives ${\cal N}=1$ supersymmetry,
especially after explicit examples of the same (and Spin(7)) 
in \cite{J,C}, exceptional holonomy compactifications of $M$-theory becomes
quite relevant for the above purpose.
In the literature, so far, the ${\cal N}=1, D=4$ Heterotic/$M$-theory dual
pair constructions, stem one way or the other from the Heterotic on $T^3$
and $M$-theory on $K3$ $D=7$ duality \cite{HT,WMtheory}. 
The question is what the ${\cal N}=1$ Heterotic/$M$-theory analog of the 
Heteoric/type IIA ${\cal N}=1$ dual pair of \cite{VW} is. 
As the $D=7$ Heterotic/$M$-Theory duality is related to
the $D=6$ Heterotic/String duality (as the decompactification limit - 
see \cite{WMtheory}), it is reasonable to think that there has to be such an 
${\cal N}=1$ Heterotic/$M$-theory dual pair. Additionally, it will be  interesting
to work out an example that is able to explicitly relate an ${\cal N}=1$ Heterotic
theory to M and F theories, as opposed to examples in the literature on only
${\cal N}=1$ 
Heteorotic/type IIA or Heterotic/M-theory or Heterotic/F-theory dual pairs.
We propose that the $M$-theory
side is given by a 7-manifold of $SU(3)\times{\bf Z}_2$-holonomy of the
type $(CY\times S^1)/g.{\cal I}$, where $g$ is a suitably defined 
freely-acting  antiholomorphic involution on the $CY$ which is precisely the
same as the one considered in \cite{VW}, $\Omega$ is the world-sheet parity
and ${\cal I}$ reflects the $S^1$. These 7-manifolds are referred to as
``barely $G_2$ manifolds'' in \cite{harvmoore}. 
In addition, the $D=4,\ {\cal N}=1$ Heterotic/F-theory dual models constructed have
the following in common (as a consequence of applying fiberwise duality to
Heterotic on $T^2$ being dual to F-theory on $K3$). 
The Heterotic theory is compactified on a
$CY_3$ that is elliptically fibered over a 2-manifold $B_2$. The F-theory
dual of this Heterotic theory is constructed by considering 
an elliptically fibered Calabi-Yau 4-fold
$X_4$ that is elliptically fibered over a 3-manifold 
$B_3$. Additionally, the base $B_3$ is a ${\bf 
P}^1$-fibration 
over $B_2$ (the same one that figures on the heterotic side). 
We propose that the required Calabi-Yau 4-fold on the F-theory side 
is elliptically fibered over a trivially rationally ruled base 
given by ${\bf CP}^1\times{\cal E}$, ${\cal E}$ being the Enriques surface.
We raise an apparent puzzle regarding the derived Hodge data and the one that one might
have naively guessed based on string/M/F dualities.

We now 
construct the M-theory uplift of type IIA background of \cite{VW}.
Now, the specific ${\cal N}=1$ Heterotic/type IIA dual pair of \cite{VW}
that we will be considering in this letter is Heterotic on a $CY$ given by
${K3\times T^2\over{\bf Z}_2^E}$ and type IIA on orientifolds of
$CY$'s (the mirrors of which are) given as hypersurface of degree 
24 in ${\bf WCP}^4[1,1,2,8,12]$, the mirror duals to which, are given by:
\begin{eqnarray}
\label{eq:K3xCP1}
& &  z_1^{24}+z_2^{24}+z_3^{12}+z_4^3+z_5^2-12\alpha z_1 z_2 z_3 z_4 z_5
-2\beta z_1^6 z_2^6 z_3^6 - \gamma z_1^{12} z_2^{12}=0.\nonumber\\
& & 
\end{eqnarray}
The ${\bf Z}_2^E$ represents the Enriques involution times reflection of
the $T^2$ as considered.
and the space-time orientation reversing antiholomorphic involution
used for constructing the CY orientifold is:
\begin{equation}
\label{eq:wdieuf}
\omega:(z_1,z_2,z_3,z_4,z_5)\rightarrow({\bar z}_2,-{\bar z}_1,{\bar z}_3,
{\bar z}_4,{\bar z}_5).
\end{equation}
Another point worth keeping in mind is that under $\omega$ of 
(\ref{eq:wdieuf}), the K\"{a}hler form $J$ going over to -$J$
is only a statement in the cohomology group $H^{1,1}$. 
One can define inhomogenous  coordinates for, e.g., Y, in the $z_2\neq0$
coordinate patch:
\begin{equation}
\label{eq:coomrds}
u\equiv{z_1\over z_2};\ v\equiv{z_3\over z_2^2};\ w\equiv{z_4\over z_2^4},
\end{equation}
[using which one can solve for ${z_5\over z_2^{12}}$ from the defining
equation (\ref{eq:K3xCP1}), and hence is not included as part of the $CY$
coordinate system]. Then, one can show that 
\begin{equation}
\label{eq:transf}
J\stackrel{\omega}{\rightarrow}-J+{\cal O}\biggl({1\over|u|^{m>0}}\ {\rm or}\
g_{u{\bar u}}-{\rm independent\ terms}\biggr),
\end{equation}
such that the $-J$ and $-J+$ extra terms both belong to the same 
cohomology class of $H^{1,1}$.
As $u\in{\bf CP}^1$-base coordinate and $g_{u{\bar u}}$ gives the size of the
${\bf CP}^1$ base, in the large base-limit of \cite{VW}, $J$ under the
antiholomorphic involution $\omega$ goes over to $-J$ {\bf exactly}. Similarly,
$H^{2,1}$ goes over to $H^{1,2}$ (and $X_{2,1}\in H^{2,1}$ goes over
to $X_{1,2}\in H^{1,2}$ exactly in the large-base limit of \cite{VW})
but an element $Y^{1,1}$ of $H^{1,1}$ goes over to an element of
the cohomology class $[-Y^{1,1}]$ of $H^{1,1}$ and no statement can 
be made for large base-limit exactness like
the ones for $J$ or $\Omega$ above. 
The exact expressions for $J$ and an element of $H^{2,1}$ under
the action of $\omega$ is given in \cite{AM2}.
To summarize, we get:
\begin{eqnarray}
\label{eq:omegaprops}
& & [\omega^*(J)]=[-J];\ \omega^*(J)\stackrel{\rm large\ {\bf CP}^1}{\longrightarrow}
-J,\nonumber\\
& & [\omega^*(X)]=[{\bar X}];\ \omega^*(X)\stackrel{\rm large\ {\bf CP}^1}{\longrightarrow}{\bar X};\nonumber\\
& & [\omega^*(Y)]=[-Y],
\end{eqnarray}
where $X\in H^{2,1}(CY_3\longrightarrow_{K3}{\bf CP}^1)$ and
$Y\in H^{1,1}(CY_3\longrightarrow_{K3}{\bf CP}^1)$, and $[\ ]$ denotes
the cohomology class. The closed and co-closed calibration 3-form $\phi$ is
given by:
\begin{equation}
\label{eq:caldieuf1}
\phi=J\wedge dx +Re(e^{-{i\theta\over2}}\Omega),
\end{equation}
where $x$ is the $S^1$ coordinate, and $\Omega$ is the holomorphic 3-form
of the $CY_3(3,243)$. 
To get the spectrum for $M$-theory compactified on the `barely $G_2$
manifold' ${\cal Z}\equiv{CY\times S^1\over \omega.{\cal I}}$, 
one sees (See \cite{harvmoore}) that ${1\over2}(H^{3,0}(CY)+H^{0,3}(CY))$ corresponding
to ${1\over 2}(h^{3,0}(CY)+h^{0,3}(CY))=1$, is invariant under the ${\bf Z}_2$
involution $\omega$. Similarly, ${1\over2}(H^{2,1}(CY)+H^{1,2}(CY))$ 
corresponding to 
${1\over2}(h^{1,2}(CY)+h^{2,1}(CY))=h^{2,1}(CY)$ elements, is invariant under
the involution $\omega$. As shown in \cite{VW}, $\omega$ acts as $-1$
on $H^{1,1}(CY)$ implying that $H^{1,1}_+(CY)$, i.e., the part of $H^{1,1}(CY)$
even under $\omega$ is zero, and the part odd, $H^{1,1}_-(CY)=H^{1,1}(CY)$.
Hence, $n_V$, $n_C$ that denote
respectively the number of vector and hypermultiplets, will be given by:
\begin{eqnarray}
\label{eq:vechyperct}
& & n_V({\cal Z})=h^{1,1}_+(CY)=0,\nonumber\\
& & n_C({\cal Z})=h^{2,1}(CY)+h^{3,0}(CY)+h^{1,1}_-(CY)=243+1+3=247.
\end{eqnarray}
This one sees that the spectra associated with Heterotic on ${K3\times T^2\over
{\bf Z}_2}$, type IIA on ${CY\over\omega.\Omega}$, and  
$M$-theory on ${CY\times S^1\over\omega.{\cal I}}$ match.

We now show the possibility of finding an ${\cal N}=1$ triality
between the ${\cal N}=1$ heterotic on $CY_3(11,11)$(/type IIA on 
${CY_3(3,243)\over\omega.\Omega}$ dual pair) of Vafa-Witten, 
$M$ theory on the ``barely $G_2$ manifold" 
${CY_3(3,243)\times S^1\over\omega.{\cal I}}$ of $SU(3)\times{\bf Z}_2$
holonomy, and F-theory on an elliptically fibered $X_4$ , 
where the ``11,11" and ``3,243" denote the Hodge numbers, 
$\omega$ is an orienation-reversing
antiholomorphic involution, ${\cal I}$ reverses the $S^1$. The $X_4$ that we obtain
in this section will be obtained by assuming that the required F-theory dual must exist.
Of course, given the basic string/M/F dualities, we know that such an F-theory dual must
exist - what we show is given the same, what the geometric data of the required $X_4$
must be.

The $CY_3$ on the heterotic side that we are interested in 
is one that is obtained by a freely-acting Enriques involution acting on the $K3$ 
times a reflection of the $T^2$, in $K3\times T^2$, i.e., the Voisin-Borcea elliptically
fibered $CY_3(11,11)\equiv {K3\times T^2\over g.{\cal I}}$, where $g$ is the generator of the
Enriques involution and ${\cal I}$ reflects the $T^2$. Hence, the $B_2$
above is ${K3\over g}$. 
Now, the ${\cal N}=2$ dual pair in \cite{KV} consisted of embedding
$SU(2)\times SU(2)$ in $E_8\times E_8$ on the Heterotic side,
resulting in $E_7\times E_7$, which is then Higgsed away. All that survives from the
$T^2$ in $K3\times T^2$ are the abelian gauge fields corresponding to  $U(1)^4$.
As shown in Vafa-Witten's paper\cite{VW}, in the ${\cal N}=1$ dual pair
obtained by suitable ${\bf Z}_2$-moddings of both sides of the ${\cal N}=2$
Heterotic/type IIA dual pair, the $U(1)^4$ gets projected out so that 
there are no vector multiplets and one gets
247 ${\cal N}=1$ chiral multiplets on the Heterotic side
on $CY_3(11,11)$. We should be able to get the same spectrum 
on the F-theory side. If $r$ denotes
the rank of the unbroken gauge group in Heterotic theory, then the number of
${\cal N}=1$ chiral multiplets in F-theory is given by the formula 
(\cite{Mohri,CL}):
\begin{equation}
\label{eq:nC}
n_C={\chi(X_4)\over6}-10+h^{2,1}(X_4) - r,
\end{equation}
which excludes the $S$ modulus of the Heterotic theory. The rank $r$ in
turn is expressed as:
\begin{equation}
\label{eq:r}
r=h^{1,1}(X_4) - h^{1,1}(B_3) - 1 + h^{2,1}(B_3).
\end{equation}
For Heterotic theory on $CY_3(11,11)$, $r=0$.

The fibration structure can be summarized as:
$X_4\longrightarrow_{T^2}B_3\longrightarrow_{{\bf CP}^1}B_2\equiv{K3\over g}
\equiv{\cal E}\equiv$Enriques surface.
Given that for elliptically fibered $X_4$, 
$h^{1,1}(X_4)-h^{1,1}(B_3)-1\geq0$, $r=0$ implies that
\begin{eqnarray}
\label{eq:consts1}
& & h^{1,1}(X_4)=h^{1,1}(B_3)+1>0;\nonumber\\
& & h^{2,1}(B_3)=0.
\end{eqnarray} 

We can write the total number of Heterotic moduli 
\begin{equation}
\label{eq:hetmood}
N_{het}=h^{1,1}(Z)+h^{2,1}(Z)+n_{bundle},
\end{equation}
where the bundle moduli correspond to  an involution $\tau$
which acts trivially on the base and as reflection of the fiber
(that can always be defined on an elliptically fibered $Z$ \cite{FMW}).
It no longer can be defined as $h^1(Z,ad(V))=I+2n_o$, where the character-valued index
$I$ is given by -$\sum_{i=}^3(-)^iTr_{H^i(Z,Ad(V))}({1+\tau\over2})=
-\sum_{i=0}^3(-)^ih^i_e(Z,Ad(V))=n_e-n_o$ for no unbroken ${\cal N}=1$ gauge group,
and $e,o$ referring to even,odd respectively under the involution $\tau$. However, given
that such an involution $\tau$ exists, one can still write that
\begin{equation}
\label{eq:gaugebundle}
n_{bundle}=n_e+n_o={\cal I}+2n_o,
\end{equation}
for a suitable ``index" ${\cal I}$. We assume that at the $\tau$-invariant point,
the action of $\tau$ can be lifted to an action of the gauge bundle embedded at
the level of $K3$. This index will encode the information about
$I(K3,Ad(SU(2)\times SU(2)))$ and the Higgsing away of the $E_7\times E_7$, or
equivalently $I(K3,Ad(E_8\times E_8))$ at the
${\cal N}=2$ level, and the freely acting Enriques involution times reflection of $T^2$.
In general, one can always write the index ${\cal I}$ as $a+b\int_{\cal E}c_1^2({\cal E})
+c\int_{\cal E}c_2({\cal E})+d\int_{\cal E}c_1^2({\cal T})+e\int_{\cal E}c_2({\cal T})
+f\int_{\cal E}c_1({\cal E})\wedge c_1({\cal T})$, where $a,b,c,d,e,f$ are constants
and ${\cal T}$ is a line bundle over ${\cal E}$.
There are no non-perturbative Heterotic 5-branes in the ${\cal N}=1$ model of \cite{VW}.
Hence, for the ${\cal N}=1$ Heterotic/F-theory duality to hold, there will no F-theory
3-branes(given by ${\chi(X_4)\over24}$) either, which implies that 
the elliptically fibered Calabi-Yau 4-fold must 
satisfy the constraint:
\begin{equation}
\label{eq:ENX_4}
\chi(X_4)=0.
\end{equation}
Assuming only a single section of the elliptic fibration: 
$Z\rightarrow_{T^2}{\cal E}(\equiv$ Enriques surface) and no 4-flux,
from general considerations (See \cite{AC}), the Hodge data of $X_4$ will be given by:
\begin{eqnarray}
\label{eq:HodgeX4}
& & h^{1,1}(X_4)=h^{1,1}(Z)+1+r=12-\int_{{\cal E}}c_1^2({\cal E}) + r,\nonumber\\
& & h^{2,1}(X_4)=n_o,\nonumber\\
& & h^{3,1}(X_4)=h^{2,1}+{\cal I}+n_o+1=12+29\int_{{\cal E}}c_1^2({\cal E})+{\cal I}
+h^{2,1}(X_4).
\end{eqnarray}
Now $t\equiv c_1({\cal T})$ (${\cal T}$ being 
a line bundle over $B_2$), the analog of $n$ in the Hirzebruch surface $F_n$, 
is a measure of the non-triviality of the ${\bf CP}^1$-fibration of the rationally ruled $B_3$.
Now, the $CY_3(3,243)$ on the type IIA side, 
can be represented as elliptic fibration over the
Hirzebruch surface $F_n$, where $n$ denotes the non-triviality of fibration
of ${\bf CP}^1_f$ over ${\bf CP}^1_b$. The Weierstrass equation for $n=0$
is given by:
\begin{equation}
\label{eq:F0}
y^2=x^3+\sum_{i=0}^8f^{(8)}_i(z_1)z_2^ix+\sum_{i=0}^{12}g^{(12)}_i(z_1)z_2^i,
\end{equation}
implying that the number of complex structure deformations, $h^{2,1}$ is
given by $9\times9+13\times13-(3+3+1)=243$. 
\footnote{Interestingly, for $n=2$, the
Weierstrass equation is given by:
\begin{equation}
\label{eq:F2}
y^2=x^3+\sum_{i=-4}^4f_{8-4i}(z_1)z_2^{4-i}x+\sum_{i=-8}^8g_{12-2i}(z_1)
z_2^{8-i},
\end{equation}
implying that the number of complex structure deformations, $h^{2,1}$ is
given by $(17+15+13+...+3+1=)81+(25+23+...+3+1=)169-(3+3+1)=243$. Hence,
elliptic fibrations over both $F_0$ and $F_2$ give the same hodge numbers.
We will work with $F_0$.}
Hence, analogous to setting $n=0$, we can  set $t=0$ and doing so would
imply the triviality of the 
fibration: $B_3={\bf CP}^1\times B_2={\bf CP}^1\times{\cal E}$,
for which $h^{2,1}(B_3)=0$ thereby satisfying (\ref{eq:consts1}).

Equating $n_{het}$ to 246, one gets from (\ref{eq:hetmood}) and (\ref{eq:gaugebundle})
the following
${\cal I}+2n_o=224$.
There are no vector multiplets, and in addition to the
heterotic dilaton, $n_{het}$ has to correspond to the number of ${\cal N}=1$
chiral multiplets $n_C$ on the F-theory side. Given that $r=\chi(X_4)=0$, from 
(\ref{eq:nC}) one gets:
$h^{2,1}(X_4)=128=n_o$.
This  gives
${\cal I}=-32$.
Using the relation: 
${\chi(X_4)\over6}=8+h^{1,1}(X_4)-h^{2,1}(X_4)+ h^{3,1}(X_4),$
one sees that the elliptically 4-manifold $X_4$ that we are
looking for is characterized by:
\begin{equation}
h^{1,1}(X_4)=12, h^{2,1}(X_4)=128, h^{3,1}(X_4)=108.
\end{equation}
This is consistent with (\ref{eq:HodgeX4}).
The $h^{2,2}(X_4)$ can be determined by the following relation
\cite{Klemmetal}
$h^{2,2}(X_4)=2(22+2h^{1,1}(X_4)+2h^{3,1}(X_4)-h^{2,1}(X_4))=268,$
which has been obtained from the definitions of elliptic
genera in terms of hodge numbers and as integrals involving
suitable powers of suitable Chern classes, and $c_1(X_4)=0$. 
Hence, ${\cal N}=1$ Heterotic Theory on ${K3\times T^2\over{\bf Z_2}}$ is dual to
F-theory on an elliptically fibered Calabi-Yau 
4-fold: $X_4[h^{1,1}=12,h^{2,1}=128,h^{3,1}=108;0]\rightarrow_{T^2}
{\bf CP}^1\times{\cal E}$. We now discuss an apparent puzzle.
At the ${\cal N}=2$ level,  Heterotic on $K3\times T^2$ should be
dual  to F-theory on $CY_3(3,243)\times T^2$ as a consequence of repeated 
fiberwise application of duality to the basic duality that Heterotic on $T^2$ is dual to
F-theory on $K3$, as well as because type IIA on a $CY_3$ should
be  dual to F-theory on $CY_3\times T^2$ and Heterotic  on $K3\times T^2$ 
is dual to type IIA on  $CY_3(3,243)$. Hence, it is possible that an orbifold 
of $K3\times T^2$ on the Heterotic side should correspond to a suitable orbifold of 
$CY_3\times T^2$ on the F-theory side. Note, however, even though a naive freely acting
orbifold of $CY_3(3,243)\times T^2$ gives the right null Euler Characteristic,
it can not, for instance, give $h^{1,1}=12$, i.e., an enhancement over the
$h^{1,1}(CY_3(3,243)\times T^2)=3+1=4$. This is unlike the case of the F-theory
dual of Heterotic on Voisin-Borcea $CY_3(19,19)$ which corresponded to an involution
with fixed points, considered in \cite{CL}. Of course, given the string/M/F dualities,
the $X_4$ with the derived fibration structure and Hodge data must exist, as 
the F-theory dual corresponding to Heterotic on $CY_3(11,11)$ must exist. One needs to
look further into this issue.
 
The $CY_4$ with the required fibration structure and Hodge data given in 
as derived above is missing from 
the list of hypersurfaces in ${\bf WCP}^5$ of Kreuzer and Skarke
because it is not possible to get the desired $CY_4$ as a 
hypersurface in any toric variety as fibrations of toric hypersurfaces
have bases that are toric varieties, and the Enriques surface, ${\cal E}$,
is not a toric variety. Perhaps, one needs a ``nef partition" 
(one could use ``nef.x" part of the package PALP\cite{KS}) 
that makes the base, ${\bf CP}^1\times{\cal E}$ a toric hypersurface. One
might have to work with complete intersections in toric varieties.

\subsection
{Type IIA on $CY_3(3,243)$ and $D=11$ Supergravity Uplift of its Orientifold} 

The periods are the building blocks, e.g., for getting the prepotential
in ${\cal N}=2$ type II theories compactified on a Calabi-Yau. 
It is in this regard that the Picard-Fuchs equation satisfied by the periods, become quite 
important. In \cite{AM2}, we addressed the issue of deriving the Picard-Fuchs equation
on the mirror Landau-Ginsburg side corresponding to the gauged linear sigma model for
a compact Calabi-Yau $CY_3(3,243)$, expressed as a degree-24 Fermat hypersurface in a 
suitableweighted complex projective space, but staying away from the orbifold singularities
by taking the large-base limit of the compact Calabi-Yau. Even though, one ended up
with more than the required number of solutions, but the essential idea that was highlighted
was the ease with which, both the large and small complex structure limits could be
addressed, and the fact that the nonanalytic $ln$-terms in the periods, could  be
easily obtained without having  to resort to parametric differentiations of infinite
series. In this paper, we address the problem of getting the right  number of the
right kind of solutions on the mirror Landau-Ginsburg side, but this time after having
resolved  the orbifold singularities. We also address the problems of showing that
unoriented instantons do not generate a superpotential on the type $IIA$ side 
in the  ${\cal N}=1$ Heterotic/type $IIA$ dual pair of \cite{VW}, whose $M$ and
$F$ theory uplifts were discussed in \cite{AM1}. It was  shown in \cite{AM2}, using
mirror symmetry, that as expected from the Heterotic and $F$ theory duals, there is
indeed no superpotential generated from ${\bf RP}^2$-instantons in the type $IIA$ side
in the large-base limit of $CY_3(3,243)$, away from the aforementioned orbifold
singularities of the relevant Fermat hypersurface. In this paper, we show that the
same remains true even after the resolution of  the orbifold singularities. Further,
we discuss the supergravity uplift of the type $IIA$ orientifold that figures in
the abovementioned ${\cal N}=1$ Heterotic/type $IIA$ dual pair, to $D=11$ supergravity.
We evaluate the K\"{a}hler potential in the large volume limit of $CY_3(3,243)$.
As an interesting aside, we give a conjecture about the action of the antiholomorphic
involution that figures in the definition  of  the type  $IIA$ orientifold, on the
periods, given its action on the cohomology of $CY_3(3,243)$,  using a canonical
(co)homology basis to expand  the holomorphic 3-form. We verify the conjecture for
$T^6$ and (partly) for the mirror  to the  quintic.  

 By following the alternative formulation of Hori and Vafa \cite{HV} for 
deriving the Picard-Fuchs equation for a definition of period integral in the mirror
Landau-Ginsburg model,
we obtain solutions valid in the large {\it and} small complex structure limits,
and get the ln terms as naturally as the analytic terms
(i.e. without using
parametric differentiation of infinite series). We also study in detail, the monodromy
matrix in the large and small complex structure limits.

Consider the Calabi-Yau 3-fold given as a degree-24 Fermat 
hypersurface in the weighted projective
space ${\bf WCP}^4[1,1,2,8,12]$:
\begin{equation}
\label{eq:Fermatdieuf}
P=z_1^{24}+z_2^{24}+z_3^{12}+z_4^{3}+z_5^2=0.
\end{equation}
It has a ${\bf Z}_2$-singularity curve and a ${\bf Z}_4$-singularity point. 
${\bf Z}_2$ and ${\bf Z}_4$ singularity resolution $\leftrightarrow$
The two new chiral superfields
 needed to be introduced as a consequence of singularity resolution,
correspond to the two ${\bf CP}^1$'s
that required to be introduced in blowing up the singularities. One then has to consider three instead
of a single $C^*$ action, and the $CY_3(3,243)$
\footnote{The $CY_3(3,243)$ considered in this paper will be
 an elliptic fibration over the Hirzebruch surface $F_2$.}
can be expressed as a suitable
holomorphic quotient corresponding to a smooth toric variety. To be more specific, one
considers the resolved Calabi-Yau $CY_3(3,243)$ as the holomorphic quotient:
${C^7-F\over(C^*)^3}|_{\rm
hyp\ constraint}$, where the diagonal $(C^*)^3$ actions
on the seven coordinates of $C^7$ are given by:
\begin{equation}
\label{eq:3Q's}
x^j\sim \lambda^{iQ^a_j}x^j,\ {\rm no\ sum\ over}\ j;\ a=1,2,3,
\end{equation}
where the three sets of charges $\{Q^{a=1,2,3}_{i=(0,),1,...,7}\}$ (the "0" being for
the extra chiral superfield with $Q^0_i=-\sum_{i=1}^7Q^a_i$\cite{W}) are 
given by the following:
\begin{equation}
\label{eq:3C*'s}
\begin{array}{ccccccccc}\\
& {\cal X}_0 &{\cal X}_1 & {\cal  X}_2 & {\cal  X}_3 & {\cal X}_4 & {\cal X}_5
& {\cal X}_6 & {\cal X}_7\\ \hline
Q^{(1)}_i:& 0 & 1 & 1 & -2
& 0 & 0 & 0 & 0 \\
Q^{(2)}_i:& 0 & 0 & 0 & 1 & 1 & 0 & 0 & -2 \\
Q^{(3)}_i: & -6 & 0 & 0 & 0 & 0 & 2 & 3 & 1 \\
\end{array}\end{equation}
where on noting:
\begin{equation}
\label{eq:linrel}
Q^{(1)}+2Q^{(2)}+4Q^{(3)}=\begin{array}{cccccccc}
-24&1&1&0&2&8&12&0
\end{array},
\end{equation}
one identifies ${\cal X}_{3,7}$ as the two extra chiral superfields introduced as a consequence of
singularity resolution.

The Landau-Ginsburg Period for the resolved $CY_3(3,243)$, as per the prescription of
Hori and Vafa, is given by: 
\begin{eqnarray}
\label{eq:PFcomp1}
& & \Pi(t_1,t_2,t_3)=\int\prod_{i=0}^7dY_i\prod_{a=1}^3dF^{(a)}
\sum_{a=1}^3d_{1a}F^{(a)}
e^{-\sum_{a=1}^3F^{(a)}(\sum_{i=1}^7Q^{(a)}_iY_i-Q^{(a)}_0Y_0-t_{(a)})-\sum_{i=0}^7e^{-Y_i}}
\nonumber\\
& & 
=\sum_{a=1}^3d_{1a}{\partial\over\partial t_{(a)}}\int\prod_{i=0}^7 dY_i\prod_{a=1}^3
\delta(\sum_{i=1}^7Q^{(a)}_iY_i-Q^{(a)}_0Y_0-t_{(a)})e^{-\sum_{i=0}^7e^{-Y_i}}\nonumber\\
& & 
\equiv\sum_{a=1}^3d_{1a}{\partial\over\partial t_{(a)}}\tilde{\Pi}(t_{1,2,3});
\end{eqnarray}
$d_{\alpha a}\equiv$ charge matrix,  $\alpha$ indexes the number of hypersurfaces
and $a$ indexes the number
of $U(1)$'s. For $CY_3(3,243)$, $\alpha=1$, $a=1,2,3$ with $d_{11}=d_{12}=0,d_{13}=6$.

Consider: 
\begin{equation}
\label{eq:dieufPi}
\tilde{\Pi}(t_{1,2,3},\{\mu_i\})\equiv\int\prod_{i=0}^7dY_i\prod_{a=1}^3
\delta(\sum_{i=1}^7Q^{(a)}_iY_i-Q^{(a)}_0Y_0-t_{(a)})e^{-\sum_{i=0}^7\mu_ie^{-Y_i}}.
\end{equation}
One can show that: 
\begin{equation}
\label{eq:redieufts}
\tilde{\Pi}(t_{1,2,3},\{\mu_i\})=\tilde{\Pi}(t_{1,2,3}^\prime,\{\mu_i=1\}),
\end{equation}
where 
\begin{equation}
\label{eq:t'sdieufs}
t_1^\prime\equiv t_1 + ln(\mu_3^2/\mu_1\mu_2),\ t_2^\prime=t_2+ln(\mu_7^2/\mu_3\mu_4),\
t_3^\prime=t_3+ln(\mu_0^6/\mu_7\mu_5^2\mu_6^3).
\end{equation}
Eliminating $Y_{0,3,7}$
 gives a order-24
Picard-Fuchs equation: 
\begin{equation}
\label{eq:PF}
{\partial^{24}\over\partial\mu_1\partial\mu_2\partial\mu_4^2
\partial\mu_5^8
\partial\mu_6^{12}}
\tilde{\Pi}(t_{1,3,4})=e^{-t_1+2t_2+4t_3}{\partial^{24}\over\partial\mu_0^{24}}
\tilde{\Pi} (t_{1,2,3}),
\end{equation}
which is the same as the PF equation for the unresolved hypersurface away from the
orbifold singularities. {\bf This overcounts the number of solutions}.

The right number of solutions must be $2h^{2,1}({\rm Mirror})+2=2.3+2=8$. To get this 
number, one notes that by adding the three constraints: 
\begin{equation}
\label{eq:consts}
Y_1+Y_2-2Y_3=t_1;\ Y_3+Y_4-2Y_7=t_2;\
-6Y_0+2Y_5+3Y_6+Y_7=t_3, 
\end{equation}
one gets: 
\begin{equation}
\label{eq:sumconsts}
-6Y_0-Y_3-Y_7+Y_1+Y_2+Y_4+2Y_5+3Y_6=t_1+t_2+t_3,
\end{equation}
which allows one to write the following order-8 PF equation:
\begin{equation}
\label{eq:PFcorrect1}
{\partial^8\over\partial\mu_1\partial\mu_2\partial\mu_4\partial\mu_5^2\partial\mu_6^3}
\tilde{\Pi}(t_{1,2,3})
=e^{-(t_1+t_2+t_3)}{\partial^8\over\partial\mu_0^6\partial\mu_3\partial\mu_7}
\tilde{\Pi}(t_{1,2,3}).
\end{equation}

If $\Theta_i\equiv{\partial\over\partial t_i^\prime}$, then one gets:
\begin{eqnarray}
\label{eq:PFcorrect2}
& & \biggl[\Theta_1^2\Theta_2\prod_{l=2}^3\prod_{k=0}^{l-1}(-l\Theta_3-k)-e^{-(t_1^\prime
+t_2^\prime+t_3^\prime)}(2\Theta_2-\Theta_3)(2\Theta_1-\Theta_2)\prod_{j=0}^5(6\Theta_3-j)
\biggr]\tilde{\Pi}=0\nonumber\\
& & {\rm with}\ z\equiv e^{-(t_1^\prime
+t_2^\prime+t_3^\prime)};\ z{d\over dz}\equiv\Delta_z,\ {\rm and\ rescaling}:
\nonumber\\
& & \biggl[\Delta_z^4(\Delta_z-{1\over2})\Delta_z(\Delta_z-{1\over3})(\Delta_z-{2\over3})
+z\prod_{j=0}^5(\Delta_z+{j\over6})\Delta_z^2\biggr]\tilde{\Pi}=0.
\end{eqnarray}

One solution to the above equation is: $\ _8F_7\left(\begin{array}{cccccccc}
0&{1\over6}&{2\over6}&{3\over6}&{4\over6}&{5\over6}&0&0\\
1&1&1&{1\over2}&1&{2\over3}&{1\over3}&\\
\end{array}\right)(-z)$

$\Biggl[$
For $e^{-t^\prime}\equiv z$, $t^\prime\equiv t_1^\prime+2t_2^\prime+4t_3^\prime$ and
suitable rescaling of $z$, the relevant order-24 PF equation {\bf for the unresolved
hypersurface} is:
\begin{equation}
\label{eq:unresPF}
\Delta^2_z\Delta_z(\Delta_z-{1\over2})\prod_{j=1}^8(\Delta_z-{j-1\over8})\prod_{j=1}^{12}
(\Delta_z-{j-1\over12})\tilde{\Pi}
=z\prod_{j=1}^{24}(\Delta_z+{j-1\over24})\tilde{\Pi}.
\end{equation}
One solution can be written in terms of the following generalized
hypergeometric function
\begin{equation}
\label{eq:PFsol1}
\ _{24}F_{23}\left(\begin{array}{ccccccc}
0 & {1\over24} & {2\over24} & {3\over24} & {4\over24} & {5\over24} & ....\ {23\over24}\\
1 & 1 & {1\over2} & {5\over8} & ...\ -{2\over8} & {1\over12} & ...\ -{10\over12}\\
\end{array}\right).
\end{equation}
$\Biggr]$

From the above solution, Meijer basis obtained using properties of 
$\ _pF_q$ and the Meijer function $I$:
\begin{eqnarray}
\label{eq:IpFqprops}
& & \ _pF_q\left(\begin{array}{cccc}
\alpha_1 & \alpha_2 & \alpha_3 & ....\ \alpha_p\\
\beta_1 & \beta_2 & \beta_3 & ....\ \beta_q\\
\end{array}\right)(z)={\prod_{i=1}^p\Gamma(\beta_i)\over\prod_{j=1}^q\Gamma(\alpha_j)}
I\left(\begin{array}{c|c}
0 & \alpha_1...\alpha_p\\
\hline
. & \beta_1...\beta_q\\
\end{array}\right)(-z)\ {\rm where}\nonumber\\
& & I\left(\begin{array}{c|c}
a_1...a_A & b_1...b_B\\
\hline
c_1...c_C & d_1...d_D\\
\end{array}\right)(z), I\left(\begin{array}{c|c}
a_1...(1-d_l)...a_A & b_1...b_B\\
\hline
c_1...c_C & d_1...\hat{d}_l...d_D\\
\end{array}\right)(-z)\nonumber\\
& & I\left(\begin{array}{c|c}
a_1...a_A & b_1..\hat{b}_j...b_B\\
\hline
c_1..(1-b_j)..c_C & d_1...d_D\\
\end{array}\right)(-z)
\end{eqnarray}
satisfy the same equation.

Now,
$z\equiv e^{-(t_1^\prime+t_2^\prime+t_3^\prime)}=e^{-(t_1+t_2+t_3)}
{\mu_1\mu_2\mu_4\mu_5^2\mu_6^3\over
\mu_0^{6}\mu_3\mu_7}.$
Hence, one can solve for large ($\equiv|z|<<1$) and small complex structure ($\equiv|z|>>1$)
limits, as well as large-size-Calabi-Yau
limit($\equiv t_i\rightarrow\infty\Leftrightarrow|z|<<1$)
on the mirror Landau-Ginsburg side with equal ease using 
Mellin-Barnes integral represention for the Meijer's function $I$,
as in \cite{GL} and in (\ref{eq:MBdieuf}) below.

Now, to get an infinite series expansion in $z$ for $|z|<1$ as well as $|z|>1$, 
one uses the following
\begin{equation}
\label{eq:MBdieuf}
I\left(\begin{array}{c|c}
a_1...a_A & b_1...b_B\\ \hline
c_1...c_C & d_1...d_D\\
\end{array}\right)(z)={1\over2\pi i}\int_\gamma ds {\prod_{i=1}^A\Gamma(a_i-s)\prod_{j=1}^B
\Gamma(b_j+s)\over\prod_{k=1}^C\Gamma(c_k-s)\prod_{l=1}^D\Gamma(d_l+s)}z^s,
\end{equation}
where the contour $\gamma$ lies to the right 
of:$s+b_j=-m\in{\bf Z}^-\cup\{0\}$ and
to the left of: $a_i-s=-m\in{\bf Z}^-\cup\{0\}$. 

This, $|z|<<1$ and $|z|>>1$ can be dealt with equal ease by suitable deformations of the
contour $\gamma$ (see Fig. 1)
to $\gamma^\prime$ and $\gamma^{\prime\prime}$ respectively (See Fig. 2). 
Additionally, instead of performing parametric differentiation of 
infinite series to get the  $ln$-terms, one get the same (for the large complex structure
limit: $|z|<1$) by evaluation of the residue at $s=0$ in the Mellin-Barnes contour integral
in (\ref{eq:MBdieuf}) as is done explicitly to evaluate the eight integrals in (\ref{eq:sols}).

The guiding principle is that of the eight solutions to $\tilde{\Pi}$, one should 
generate solutions in which one gets $(ln z)^P$, $P=1,...,4$ so that one gets 
$(ln z)^{P-1}$ for $\Pi$, and one can then identify terms independent of $ln z$ with
$Z^0$, three $(ln z)$ terms with $Z^{1,2,3}$, three $(ln z)^{P\leq2}$ 
terms with $F_{1,2,3}\equiv
{\partial F\over\partial Z^{1,2,3}}$,
and finally $(ln z)^{P\leq3}$ term with $F_0\equiv{\partial F\over\partial Z^0}$.

One (non-unique) choice of solutions for $\tilde{\Pi}(z)$ is given below:
\begin{eqnarray}
\label{eq:sols}
z{d\over dz}\left[\begin{array}{c}
I\left(\begin{array}{c|c}
0\ 0 & 0\ {1\over6}\ {2\over6}\ {3\over6}\ {4\over6}\ {5\over6}\ 0\ 0 \\
& \\ \hline
&\\
. & 1\ 1\ {1\over2}\ 1\ {2\over3}\ {1\over3}\\
\end{array}\right)(-z)\\ 
\\
I\left(\begin{array}{c|c}
0\ 0 {1\over2}\ {1\over3}\ {2\over3} & 
{1\over6}\ {2\over6}\ {3\over6}\ {4\over6}\ {5\over6}\\
& \\ \hline
&\\
1\ 1\ 1 & 1\ 1\ 1\\
\end{array}\right)(-z)\\
\\
I\left(\begin{array}{c|c}
0\ 0 \ {1\over2}& 0\ {1\over6}\ {2\over6}\ {3\over6}\ {4\over6}\ {5\over6}\ 0\ \\
& \\ \hline
&\\
1 & 1\ 1\ 1\ {2\over3}\ {1\over3}\\
\end{array}\right)(-z)\\
\\
I\left(\begin{array}{c|c}
0\ 0 \ {1\over3}& 0\ {1\over6}\ {2\over6}\ {3\over6}\ {4\over6}\ {5\over6}\ 0\ \\
& \\ \hline
&\\
1 & 1\ 1\ {1\over2}\ 1\ {1\over3}\\
\end{array}\right)(z)\\
\\
I\left(\begin{array}{c|c}
0\ 0 \ {2\over3}& 0\ {1\over6}\ {2\over6}\ {3\over6}\ {4\over6}\ {5\over6}\ 0\ \\
& \\ \hline
&\\
1 & 1\ 1\ {1\over2}\ 1\ {2\over3}\\
\end{array}\right)(-z)\\
\\
I\left(\begin{array}{c|c}
0\ 0 \ {1\over2}& {1\over6}\ {2\over6}\ {3\over6}\ {4\over6}\ {5\over6}\ 0\ \\
& \\ \hline
&\\
1\ 1 & 1\ 1\ 1\ {2\over3}\ {1\over3}\\
\end{array}\right)(z)\\
\\
I\left(\begin{array}{c|c}
0\ 0 \ {1\over3}& {1\over6}\ {2\over6}\ {3\over6}\ {4\over6}\ {5\over6}\ 0\ \\
& \\ \hline
&\\
1\ 1 & 1\ 1\ {1\over2}\ {2\over3}\ {1\over3}\\
\end{array}\right)(z)\\
\\
I\left(\begin{array}{c|c}
0\ 0 \ {2\over3}& {1\over6}\ {2\over6}\ {3\over6}\ {4\over6}\ {5\over6}\ 0\ \\
& \\ \hline
&\\
1\ 1 & 1\ 1\ {1\over2}\ {2\over3}\\
\end{array}\right)(z)\\
\end{array}\right]\sim\left[\begin{array}{c}
F_0\\
Z^0\\
F_1\\
F_2\\
F_3\\
Z^1\\
Z^2\\
Z^3\\
\end{array}\right]
\end{eqnarray}

(a)
\begin{eqnarray}
\label{eq:reltoF_0}
& & I\left(\begin{array}{c|c}
0\ 0 & 0\ {1\over6}\ {2\over6}\ {3\over6}\ {4\over6}\ {5\over6}\ 0\ 0 \\
& \\ \hline
& \\
. & 1\ 1\ {1\over2}\ 1\ {2\over3}\ {1\over3}\\
\end{array}\right)(-z)={1\over2\pi i}\int_\gamma ds{[\Gamma(-s)]^2[\Gamma(s)]^3\prod_{j=1}^5
\Gamma(s+{j\over6})\over[\Gamma(s+1)]^3\Gamma({1\over2}+s)\Gamma({2\over3}+
s)\Gamma({1\over3}+s)} (-z)^s
\nonumber\\
& & =\theta(1-|z|)\Biggl[{2(2\pi)^{{3\over2}}\over\sqrt{\pi}}\biggl[\biggl(ln\biggl({2^23^3\over6^6}\biggr) + ln(-z)\biggr)^4+{65\over3}
\biggl(ln\biggl({2^23^3\over6^6}\biggr) + ln(-z)\biggr)^2\nonumber\\
& & +{169\pi^4\over3}-1440\zeta(3)\biggl(
ln\biggl({2^23^3\over6^6}\biggr)+ln(-z)\biggr)+402\zeta(4)\biggr]
+\sum_{m=1}^\infty{(2\pi)^{{3\over2}}2^{{1\over2}-6m}\Gamma(6m)\over m^3(m!)^23^{3m}
\Gamma({1\over2}+m)\Gamma(3m)}\nonumber\\
& & \times\biggl[2\gamma+6\Psi(6m)-2\Psi(2m)-3\Psi(m)-3m^2+
ln\biggl({2^23^3\over6^6}\biggr)\biggr](-z)^m\Biggr]\nonumber\\
& & -\theta(|z|-1)\Biggl[ \sum_{m=0}^\infty{(-)^m
(\Gamma(m+{1\over6})^2\Gamma(-m+{2\over3})\over(m+{1\over6})^3m!}(-z)^{-m-{1\over6}}+
\nonumber\\
& & 
\sum_{m=0}^\infty{(-)^m(\Gamma(m+{5\over6})^2\Gamma(-m-{2\over3})\over(m+{5\over6})^3m!}(-z)^{-m-{5\over6}}\Biggr]
\end{eqnarray}

The connection between (\ref{eq:sols}) 
that {\it effectively depends
only on one complex structure parameter
$z=e^{-(t_1+t_2+t_3)}{1\over z_1z_2z_3}$}, and the solutions
 given in the literature \cite{HKTY}of the form:
\begin{equation}
\label{eq:solklemm}
\partial_{\rho_m}^{s_m}\partial_{\rho_n}^{s_n}\partial_{\rho_p}^{s_p}
\sum_{m,n,p}
c(m,n,p;\rho_m,\rho_n,\rho_p)
z_1^{m+\rho_m}z_2^{n+\rho_n}z_3^{p+\rho_p}|_{\rho_m=\rho_n=\rho_p=0},
\end{equation}
with $s_m+s_n+s_p\leq3$,
and $z_1\equiv{\mu_1\mu_2\over\mu_3^2},\ z_2\equiv{\mu_3\mu_4\over\mu_7^2},\ z_3\equiv
{\mu_7\mu_5^2\mu_6^3\over\mu_0^6}$, needs to be understood.
The appearance of $\partial_{\rho_m}^{s_m}\partial_{\rho_n}^{s_n}\partial_{\rho_p}^{s_p}
\sum_{m,n,p}$ in (\ref{eq:solklemm}) is what was referred to earlier on as
parametric differentiation of infinite series, something which, as we have explicitly shown
above, is not needed in the approach followed in this work.

The Picard-Fuchs equation can be written in the form\cite{M}:
\begin{equation}
\label{eq:PFform}
\biggl(\Delta_z^{8}+\sum_{i=1}^{7}{\bf B}_i(z)\Delta_z^i\biggr)\tilde{\Pi}(z)
=0.
\end{equation}
The Picard-Fuchs
 equation in the form written in (\ref{eq:PFform}) can alternatively be expressed as
the following system of eight linear differential equations:
\begin{eqnarray}
\label{eq:PFdiffeq3}
& & \Delta_z\left(\begin{array}{c}\\
\tilde{\Pi}(z)\\
\Delta_z
\tilde{\Pi}(z)\\
(\Delta_z)^2
\tilde{\Pi}(z)\\
...\\
(\Delta_z)^7
\tilde{\Pi}(z)\\
\end{array}\right)
=\nonumber\\
& & \left(\begin{array}{ccccc}\\
0 & 1 & 0 & ... 0 & 0\\
0 & 0 & 1 & ... 0 & 0\\
. & . & . & ... . & . \\
0 & 0 & 0 & ... 0 & 1 \\
0 & -{\bf B}_1(z) & -{\bf B}_2(z) & ... -{\bf B}_6(z) & -{\bf B}_7(z)\\
\end{array}\right)
\left(\begin{array}{c}\\
\tilde{\Pi}(z)\\
\Delta_z
\tilde{\Pi}(z)\\
(\Delta_z)^2
\tilde{\Pi}(z)\\
...\\
(\Delta_z)^7
\tilde{\Pi}(z)\\
\end{array}\right)
\end{eqnarray}
The matrix on the RHS of (\ref{eq:PFdiffeq3}) is usually denoted by $A(z)$.
 
If the eight solutions, $\{\tilde{\Pi}_{I=1,...,8}\}$,
 are collected as a column vector $\tilde{\Pi}(z)$, then the {\it constant}\footnote{This 
thus implies that both $\tilde{\Pi}$ and $\Pi$, have the {\it same} monodromy matrix.}
monodromy matrix $T$ for $|z|<<1$ is defined by:
\begin{equation}
\tilde{\Pi}(e^{2\pi i}z)=T\tilde{\Pi}(z).
\end{equation}
The basis for the space of solutions can be collected as the
columns of the ``fundamental matrix" $\Phi(z)$ given by:
\begin{equation}
\label{eq:PFdiffeqsol}
\Phi(z)=
S_8(z)z^{R_8},
\end{equation}
where $S_8(z)$ and $R_8$ are 8$\times$8 matrices that single and
multiple-valued respectively. Note that ${\bf B}_i(0)\neq0$, which influences
the monodromy properties. Also,
\begin{equation}
\label{eq:Phidieuf2}
\Phi(z)_{ij}=\left(\begin{array}{ccc}\\
\tilde{\Pi}_1(z) & ... & \tilde{\Pi}_8(z) \\
\Delta_z \tilde{\Pi}_1(z) & ... & \Delta \tilde{\Pi}_8(z) \\
\Delta_z^2 \tilde{\Pi}_2(z) & ... & \Delta^2 \tilde{\Pi}_8(z) \\
... & ... & ... \\
\Delta_z^7 \tilde{\Pi}_1(z) & ... & \Delta_z^7 \tilde{\Pi}_8(z)\\
\end{array}\right)_{ij},
\end{equation}
implying that
\begin{equation}
\label{eq:Tdieuf}
T=e^{2\pi i R^t}.
\end{equation}
Now, writing $z^R=e^{R ln z}=1 + R ln z + R^2(ln z)^2+...$, and further
noting that there are no terms of order higher than $(ln z)^4$ in
$\tilde{\Pi}(z)$
obtained above, implies that the matrix $R$ must satisfy the property:
$R^{m}=0,\ m=5,...\infty$. Hence, $T=e^{2\pi i R^t}=1 + 2\pi i R^t +
{(2\pi i)^2\over2}(R^t)^2+{(2\pi i)^3\over 6}(R^t)^3+{(2\pi i)^4\over24}(R^t)^4.$ 
Irrespective of whether or not the
distinct eigenvalues of $A(0)$ differ by integers, one has to evaluate
$e^{2\pi i A(0)}$. The eigenvalues of $A(0)$ of (\ref{eq:A(0)}), are $0^5,{1\over3},{1\over2},
{2\over3}$, and hence five of the eight eigenvalues differ by an integer (0). 

Now, the Picard-Fuchs equation (\ref{eq:PFcorrect2}) can be rewritten in the form
(\ref{eq:PFform}), with the following values of $B_i$'s:
\begin{eqnarray}
\label{eq:Bis}
& & {\bf B}_{1,2}=0, {\bf B}_3={5z\over324(1+z)}, {\bf B}_4={137\over648(1+z)}
\nonumber\\
& & {\bf B}_5={({25\over24}z-{1\over9})\over(1+z)},
 {\bf B}_6={(26+85z)\over36(1+z)}, {\bf B}_7=-{(3-5z)\over2(1+z)}.
\end{eqnarray}

Under the change of basis
$\tilde{\Pi}(z)\rightarrow \tilde{\Pi}^\prime(z)=M^{-1}\tilde{\Pi}(z)$, 
and writing $\tilde{\Pi}_j(z)=\sum_{i=0}^4(ln z)^iq_{ij}(z)$ (See \cite{GL} and the
appendix), one sees that 
\begin{eqnarray}
\label{eq:primed}
& & \tilde{\Pi}^\prime_j(z)=\sum_{i=0}^4(ln z)^iq^\prime_{ij}(z),\nonumber\\
& & q^\prime(z)=q(z)(M^{-1})^t,\nonumber\\
& &  \Phi^\prime(z)\Phi(z)(M^{-1})^t,\ S^\prime(z)=S(z)(M^{-1})^t,\ R^\prime=M^tR(M^{-1})^t.
\end{eqnarray}
By choosing $M$ such that
$S^\prime(0)={\bf 1}_{24}$, one gets 
\begin{equation}
\label{eq:monodromy1}
T(0)=M(e^{2i\pi A(0)})^tM^{-1}.
\end{equation}
The matrix $A(0)$ is given by:
\begin{equation}
\label{eq:A(0)}
A(0)=\left(\begin{array}{cccccccc}\\
0&1&0&0&0&0&0&0\\
0&0&1&0&0&0&0&0\\
0&0&0&1&0&0&0&0\\
0&0&0&0&1&0&0&0\\
0&0&0&0&0&1&0&0\\
0&0&0&0&0&0&1&0\\
0&0&0&0&0&0&0&1\\
0&0&0&0&0&{1\over9}&-{13\over18}&{3\over2}\\
\end{array}\right)
\end{equation}

Using MATHEMATICA, one then can evaluate the ``matrix exponent" involving $A(0)$.

Writing the solution vector $\tilde{\Pi}_i$ as $\tilde{\Pi}_i=\sum_{j=0}^4(ln z)^j q_{ji}$
(following the notation of \cite{GL}), one notes:
\begin{equation}
\label{eq:primedsol1}
(\Phi^\prime)_i=(\tilde{\Pi}^\prime)^t_i=\biggl(S^\prime z^{A(0)}\biggr)_{0i}=
(ln z)^j q^\prime_{ji}.
\end{equation}
From (\ref{eq:primedsol1}), one gets the following:
\begin{equation}
\label{eq:primedsol2}
(q^\prime(0))_{ji}={\delta_{ji}\over j!},\ 0\leq (i,j)\leq 4.
\end{equation}
For $5\leq i\leq 7$, consider. e.g., $i=5$. Then from the expression for $z^{A(0)}$ above,
\begin{eqnarray}
\label{eq:primedsol3}
& & \sum_{j=0}^4(q^\prime)_{j5}(ln z)^j=
(S^\prime)_{00}[f_{05}(z^{{1\over2}},z^{{1\over3}})
+\sum_{n=1}^4c_n^{(05}(ln z)^n]+
(S^\prime)_{01}[f_{15}(z^{{1\over2}},z^{{1\over3}})+\sum_{n=1}^3c_n^{(25)}(ln z)^n]\nonumber\\
& & 
(S^\prime)_{02}[f_{25}(z^{{1\over2}},z^{{1\over3}})
+\sum_{n=1}^2c_n^{(25}(ln z)^n]+
(S^\prime)_{03}[f_{35}(z^{{1\over2}},z^{{1\over3}})+c_1^{(35)}(ln z)]\nonumber\\
& & 
+(S^\prime)_{04}f_{45}(z^{{1\over2}},z^{{1\over3}})+(S^\prime)_{05}f_{55}(z^{{1\over3}})
+(S^\prime)_{06}f_{65}(z^{{1\over2}},z^{{1\over3}})+(S^\prime)_{07}f_{75}(z^{{1\over3}})
\end{eqnarray}
where the  $f_{ij}$'s and $c_n^{ij}$'s can be determined from the expression for
$z^{A(0}$ given below. From (\ref{eq:primedsol3}),
one gets:
\begin{eqnarray}
\label{eq:primedsol4}
& & (q^\prime)_{05}=\sum_{i=0}^7(S^\prime)_{0i}f_{i5},
 (q^\prime)_{15}=\sum_{i=0}^3(S^\prime)_{0i}c_1^{i5},
 (q^\prime)_{25}=\sum_{i=0}^2(S^\prime)_{0i}c_2^{i5}\nonumber\\
& & (q^\prime)_{35}=\sum_{i=0}^1(S^\prime)_{0i}c_3^{i5},
 (q^\prime)_{45}=(S^\prime)_{00}c_4^{05}.
\end{eqnarray}
From (\ref{eq:primedsol4}), one gets:
\begin{eqnarray}
\label{eq:primedsol6}
& & (q^\prime(0))_{0i}=f_{0i}(0);
(q^\prime(0))_{ij}=c_i^{0j},\ 1\leq i\leq 4,\ 5\leq j\leq 7.
\end{eqnarray}

Again using the MATHEMATICA notebook format, the value of $z^{A(0)}$, as
evaluated by MATHEMATICA is given by:
The matrix $q^\prime$ introduced in (\ref{eq:primedsol1}) is given by:
\begin{equation}
\label{eq:q'dieuf}
q^\prime=\left(\begin{array}{cccccccc}
1 & 0 & 0 & 0 & 0 & -\left( \frac{391933}{32} \right)  & \frac{130077}{32} & 
-\left( \frac{53739}{16} \right)  \cr 0 & 1 & 0 & 0 & 0 & -\left( \frac{5971}
     {16} \right)  & \frac{19215}{16} & -\left( \frac{7785}{8} \right)  \cr 0 & 0 & \frac{1}{2} & 0 & 0 & -\left( \frac{865}{16} \right)  & \frac{2637}{16} & -\left
     ( \frac{1035}{8} \right)  \cr 0 & 0 & 0 & \frac{1}{6} & 0 & -\left( \frac{115}{24} \right)  & \frac{105}{8} & -\left( \frac{39}{4} \right)
      \cr 0 & 0 & 0 & 0 & \frac{1}{24} & -\left( \frac{13}{32} \right)  & \frac{9}{16} 
& -\left( \frac{3}{8} \right)  \cr 
\end{array}\right)
\end{equation} 

Further, the matrix $q$ is of the form:
\begin{equation}
\label{eq:qdieuf}
q=\left(\begin{array}{cccccccc} 
q_{00} & q_{01} & q_{02} & q_{03} & 
q_{04} & q_{05} & q_{06} & q_{07} \cr 
   q_{10} & q_{11} & q_{12} & q_{13} & q_{14} & q_{15} & q_{16} & q_{17} \cr 0 & q_{21} & q_{22} & q_{23} & q_{24} 
& q_{25} & q_{26} & q_{27} \cr 0 & 0 & 0 & 0 & q_{34} & 
q_{35} & q_{36} & q_{37} \cr 
0 & 0 & 0 & 0 & q_{44} & 0 & 0 & 0 \cr  
\end{array}\right)
\end{equation}

From the matrix equation $q^\prime=q(M^{-1})^t$, one 
sees that one has 40 equations in 64 variables, one has the freedom to (judiciously)
give arbitrary values to 24 variables, bearing in mind that from the forms of
the matrices $q^\prime$ and $q (M^{-1})^t$, the values of $(M^{-1})^t_{4i}$ are fixed.
We set: $M_{ij}=\delta_{ij}$ for $0\leq i\leq 3$ and $0\leq j\leq 7$.

The above matrix equation can then be solved for the 64-24=40 entries $(
M^{-1})^t_{ij}$ $4\leq i\leq 7,\ 0\leq j\leq 7$ to give the following result: 

$(M^{-1})^t_{30}=\frac{X_{30}}{Y_{30}}$, where

$ X_{30}= {q_{06}}\,{q_{10}}\,{q_{27}}\,{q_{35}} + {q_{16}}\,{q_{27}}\,{q_{35}} - 
            {q_{00}}\,{q_{16}}\,{q_{27}}\,{q_{35}} - {q_{05}}\,{q_{10}}\,{q_{27}}\,{q_{36}} - 
            {q_{15}}\,{q_{27}}\,{q_{36}} + {q_{00}}\,{q_{15}}\,{q_{27}}\,{q_{36}} 
( -1 + {q_{00}} ) \,{q_{17}}\,( {q_{26}}\,{q_{35}} - {q_{25}}\,{q_{36}} )  + 
            {q_{07}}\,{q_{10}}\,( -( {q_{26}}\,{q_{35}} )  + {q_{25}}\,{q_{36}} )  -$

\noindent $ {q_{06}}\,{q_{10}}\,{q_{25}}\,{q_{37}} - {q_{16}}\,{q_{25}}\,{q_{37}} + 
            {q_{00}}\,{q_{16}}\,{q_{25}}\,{q_{37}} + {q_{05}}\,{q_{10}}\,{q_{26}}\,{q_{37}} + 
            {q_{15}}\,{q_{26}}\,{q_{37}} - {q_{00}}\,{q_{15}}\,{q_{26}}\,{q_{37}}$ 

$Y_{30}=-( {q_{03}}\,
               {q_{17}}\,{q_{26}}\,{q_{35}} )  + {q_{03}}\,{q_{16}}\,{q_{27}}\,{q_{35}} - 
            {q_{05}}\,{q_{17}}\,{q_{23}}\,{q_{36}} + {q_{03}}\,{q_{17}}\,{q_{25}}\,{q_{36}} + 
            {q_{05}}\,{q_{13}}\,{q_{27}}\,{q_{36}} -
 {q_{03}}\,{q_{15}}\,{q_{27}}\,{q_{36}} + {q_{07}}\,( -( {q_{16}}\,{q_{23}}\,{q_{35}} )  
+ {q_{13}}\,{q_{26}}\,{q_{35}} + 
               {q_{15}}\,{q_{23}}\,{q_{36}} - {q_{13}}\,{q_{25}}\,{q_{36}} )  + $

\noindent$ {q_{05}}\,{q_{16}}\,{q_{23}}\,{q_{37}} - {q_{03}}\,{q_{16}}\,{q_{25}}\,{q_{37}} - 
            {q_{05}}\,{q_{13}}\,{q_{26}}\,{q_{37}} + {q_{03}}\,{q_{15}}\,{q_{26}}\,{q_{37}} +$

\noindent$   {q_{06}}\,( {q_{17}}\,{q_{23}}\,{q_{35}} - {q_{13}}\,{q_{27}}\,{q_{35}} -\, 
{q_{15}}\,{q_{23}}\,{q_{37}} + {q_{13}}\,{q_{25}}\,{q_{37}} ) , $

\vskip 0.2in $(M^{-1})^{t}_{31}=\frac{X_{31}}{Y_{31}}$ where
 
$X_{31}={q_{01}}{q_{17}}{q_{26}}{q_{35}} - {q_{01}}{q_{16}}{q_{27}}{q_{35}} + 
            {q_{05}}{q_{17}}{q_{21}}{q_{36}} - {q_{01}}{q_{17}}{q_{25}}\,{q_{36}} + 
            {q_{05}}{q_{27}}{q_{36}} - {q_{05}}{q_{11}}{q_{27}}{q_{36}} + 
            {q_{01}}{q_{15}}{q_{27}}{q_{36}} + 
            {q_{07}}( {q_{16}}{q_{21}}{q_{35}} + 
               {q_{26}}( {q_{35}} - {q_{11}}\,{q_{35}})  - $

\noindent$               ( {q_{15}}\,{q_{21}} + {q_{25}} - {q_{11}}\,{q_{25}} ) \,{q_{36}} )  - 
            {q_{05}}\,{q_{16}}\,{q_{21}}\,{q_{37}} + {q_{01}}\,{q_{16}}\,{q_{25}}\,{q_{37}} - 
            {q_{05}}\,{q_{26}}\,{q_{37}} + {q_{05}}\,{q_{11}}\,{q_{26}}\,{q_{37}} - 
            {q_{01}}\,{q_{15}}\,{q_{26}}\,{q_{37}} + 
            {q_{06}}\,( -( {q_{17}}\,{q_{21}}\,{q_{35}} )  + 
               ( -1 + {q_{11}} ) \,{q_{27}}\,{q_{35}} +$

\noindent$( {q_{15}}\,{q_{21}} + {q_{25}} - {q_{11}}\,{q_{25}} ) \,{q_{37}} )$

$Y_{31}=-( 
               {q_{03}}\,{q_{17}}\,{q_{26}}\,{q_{35}} )  + {q_{03}}\,{q_{16}}\,{q_{27}}\,{q_{35}} - 
            {q_{05}}\,{q_{17}}\,{q_{23}}\,{q_{36}} + {q_{03}}\,{q_{17}}\,{q_{25}}\,{q_{36}} + 
            {q_{05}}\,{q_{13}}\,{q_{27}}\,{q_{36}} - {q_{03}}\,{q_{15}}\,{q_{27}}\,{q_{36}} + 
            {q_{07}}\,( -( {q_{16}}\,{q_{23}}\,{q_{35}} )  + {q_{13}}\,{q_{26}}\,{q_{35}} + 
               {q_{15}}\,{q_{23}}\,{q_{36}} - {q_{13}}\,{q_{25}}\,{q_{36}} )  + 
            {q_{05}}\,{q_{16}}\,{q_{23}}\,{q_{37}} - {q_{03}}\,{q_{16}}\,{q_{25}}\,{q_{37}} - 
            {q_{05}}\,{q_{13}}\,{q_{26}}\,{q_{37}} + {q_{03}}\,{q_{15}}\,{q_{26}}\,{q_{37}} + $

\noindent${q_{06}}\,( {q_{17}}\,{q_{23}}\,{q_{35}} - {q_{13}}\,{q_{27}}\,{q_{35}} - 
               {q_{15}}\,{q_{23}}\,{q_{37}} + {q_{13}}\,{q_{25}}\,{q_{37}} ),$

\vskip 0.2in $(M^{-1})^{t}_{40}=(M^{-1})^{t}_{41}=(M^{-1})^{t}_{42}=(M^{-1})^{t}_{43}=0,
(M^{-1})^{t}_{44}=\frac{1}{24\,{q_{44}}},$

\noindent$
(M^{-1})^{t}_{45}=\frac{-13}{32\,{q_{44}}},
(M^{-1})^{t}_{46}=\frac{9}{16\,{q_{44}}},(M^{-1})^{t}_{47}=\frac{-3}{8\,{q_{44}}}$
\vskip 0.2in
In the  above expressions for  $(M^{-1})^t$, the non-zero $q_{ij}'s,\ 0\leq i\leq 4,\ 
0\leq j\leq 7$ are given in \cite{AM3}, e.g.
\begin{eqnarray}
\label{eq:qijdieufs}
& & q_{00}={8\pi^3\over3}
\biggl[ln\biggl({2^2.3^3\over6^6}\biggr)+i\pi\biggr]\nonumber\\
& & q_{01}=2\pi^2\biggl[ln\biggl({2^2.3^3\over6^6}\biggr)+{14\over3}\pi^2\biggr]\nonumber\\
& & q_{02}={4\pi^2\over\sqrt{3}} 
\biggl(\biggl[ln\biggl({2^2.3^3\over6^6}\biggr)+{\pi\over\sqrt{3}}\biggr]^2+5\pi^2\biggr)
\nonumber\\
& & q_{03}=
\biggl(\biggl[ln\biggl({2^2.3^3\over6^6}\biggr)-{\pi\over\sqrt{3}}\biggr]^2+5\pi^2\biggr)
\nonumber\\
& & q_{04}=4\sqrt{2}\pi\Biggl(
\biggl[ln\biggl({2^2.3^3\over 6^6}\biggr)+i\pi\biggr]^4
+{65\over3}\biggl[ln\biggl({2^2.3^3\over 6^6}\biggr)+i\pi\biggr]^2+{169\pi^4\over3}\nonumber\\
& & -1440\biggl[ln\biggl({2^2.3^3\over 6^6}\biggr)+i\pi\biggr]+402\zeta(4)\Biggr]\nonumber\\
& & q_{05}=2\pi^2\Biggl(
\biggl[ln\biggl({2^2.33^3\over6^6}\biggr)+i\pi\biggr)^3
+15\pi^2ln\biggl({2^2.33^3\over6^6}\biggr)-356\zeta(3)\Biggr)\nonumber\\
\end{eqnarray}
The matrix $(M^{-1})^t$ is non-singular as the determinant is non-zero.
Using MATHEMATICA,
one can actually evaluate $T$, but the expression is extremely long and complicated
and  will not be given in this paper.

The monodromy around $z=\infty$ can be evaluated as follows(similar to the way
given in \cite{GL}). For $|z|>>1$, one can write:
\begin{equation}
\label{eq:infty1}
\tilde{\Pi}_a(z)=\sum_{j=1}^5A_{aj}(z)u_j(z),\ a=0,...,7,
\end{equation}
where $u_j(z)=e^{-{j\over6}}$. Now, as $z\rightarrow e^{2i\pi}z$, with obvious meanings
to the notation:
\begin{equation}
\label{eq:infty2}
T_u(\infty)=\left(\begin{array}{ccccc}
e^{-i{\pi\over3}}&0&0&0&0\\
0&e^{-{2i\pi\over3}}&0&0&0\\
0&0&e^{-i\pi}&0&0\\
0&0&0&e^{-{4i\pi\over3}}&0\\
0&0&0&0&e^{-{5i\pi\over3}}\\
\end{array}\right).
\end{equation}
Now, using
\begin{equation}
\label{eq:infty3}
\tilde{\Pi}(z\rightarrow e^{2i\pi}z)|_{z\rightarrow\infty}
={\cal A}(z\rightarrow e^{2i\pi}z)T_u(\infty)u(z)|_{z\rightarrow\infty}
\equiv T(\infty){\cal A}(z)u(z)
|_{z\rightarrow\infty}.
\end{equation}
So, equation (\ref{eq:infty3}) is the defining equation for the monodromy matrix 
around $z\rightarrow\infty$. Note, however, that from the point of view of computations,
given that the matrix $A$ is not a square matrix, (\ref{eq:infty3}) involves solving
40 equations in 64 variables. 
The $8\times 5$ matrix $A_{ai}(\infty)$ 
with $a=0,...,7\ i=1,...,5$ for $z\rightarrow\infty$, 
is given below: 
\begin{eqnarray}
\label{eq:Anonzero}
& & {\cal A}(\infty)=\nonumber\\
& & \left(\begin{array}{ccccc}
-(2\pi)^{{5\over2}}6^{{3\over2}}{\sqrt{\pi}\Gamma({2\over3})\over\Gamma(
{7\over6})(\Gamma({5\over6})^2} &
54\sqrt{3}\pi^{{3\over2}}{\Gamma({5\over6})\over\Gamma({4\over3})\Gamma
({2\over3})} & -288\sqrt{{\pi\over3}} & 
-3\pi^2{\Gamma({7\over6})\Gamma(-{1\over3})\over\Gamma({5\over3})
\Gamma({1\over3})} & {12\pi\over25\sqrt{3}}{\Gamma(-{2\over3})
\Gamma(-{1\over6})\over\Gamma({11\over6})\Gamma({1\over6})}\\
{216(\Gamma({2\over3})^2\Gamma({1\over3})\over(\Gamma({5\over6}))^2} & 0 
& -{16\over\sqrt{3}\pi} & 0 & -{144\sqrt{3}\pi\over125}{\Gamma(-{2\over3})\over
(\Gamma({1\over6}))^2}\\
{216\pi\Gamma({2\over3})\over(\Gamma({5\over6}))^2} & 0 & 
0 & {-27\sqrt{\pi}\Gamma({1\over6})\over4(\Gamma({1\over3})^2}
& {36\over125}\Gamma(-{2\over3})\Gamma(-{1\over6})\\
{216\Gamma({1\over6})\Gamma({2\over3})\over\Gamma({5\over6})} &
{-27\pi\Gamma(-{2\over3})\over8(\Gamma({1\over6}))^2} & 0 
& -{27\pi\Gamma(-{2\over3})\over8(\Gamma({1\over6}))^2} & 0 \\
-216(\Gamma({1\over6}))^2\Gamma({2\over3})
& 0 & 0 & 0 &-{216\over125}(\Gamma({5\over6}))^2\Gamma(-{2\over3})\\
-{432\pi\Gamma({1\over6})\over\sqrt{3}\Gamma({5\over6})} & 0 &
2\sqrt{3}\pi & 0 & 
-{216\over125}{\Gamma({5\over6})\Gamma(-{2\over3})\Gamma(-{1\over3})
\over\Gamma({1\over6})}\\
{216\pi\Gamma({1\over6})\Gamma({2\over3})\over\Gamma({5\over6})}
&0  & 0 & -{27\sqrt{\pi}\over4}{\Gamma({2\over3})\Gamma({1\over6})\over
\Gamma({1\over3})}& 
{216\over125}{\Gamma({5\over6})\Gamma({7\over6}\Gamma(-{2\over3})\Gamma
(-{1\over6})\over\Gamma({1\over6})}\\
216(\Gamma({1\over6}))^2\Gamma({2\over3})
& 27\sqrt{\pi}\Gamma({1\over3})\Gamma(-{1\over6}) & 0 & 0 &
-{216\over125}{\Gamma({5\over6})\Gamma(-{2\over3})
\over\Gamma({1\over6})}\\
\end{array}\right)
\end{eqnarray}
Given that $(T_u)_{ij}=e^{-{\sqrt{-1}\pi i\over3}}\delta_{ij}$, no sum over $i$, 
one sees that the equation
(\ref{eq:infty3}) becomes:
\begin{equation}
\label{eq:infty4}
e^{-{\sqrt{-1}\pi j\over3}}A_{aj}(\infty)=T_{ab}(\infty)A_{bj}(\infty),
\end{equation}
(no sum over $j$) which needs to be solved  for $T_{ab}(\infty)$. MATHEMATICA is unable
to perform the required computation - however, it is in principle, doable.

Unfortunately,
MATHEMATICA is not able to handle such a computation, this time. However, it is clear
that it is in prinicple, a doable computation.

As done in \cite{Witten}, consider $F$-theory on an
elliptically fibered Calabi-Yau 4-fold $X_4$ with holomorphic map 
$\pi:X_4\rightarrow B_3$ and a 6-divisor $D_3$
as a section such that $\pi(D_3)=C_2\subset B_3$. Then for vanishing size of the the elliptic
fiber, it was argued in \cite{Witten} that 5-branes 
wrapped around $D_3$ in $M$-theory on the same $X_4$ obeying
the unit-arithmetic genus condition, $\chi(D_3,{\cal O}_{D_3})=1,$
correspond to 3-branes wrapped around $C_2$ in type $IIB$, or equivalently $F$-theory 3-branes wrapped around
$C_2\subset B_3$. It was shown in \cite{Witten} that 
only 3-branes contribute to the superpotential in $F$-theory. 
As there are no 3-branes in the $F$-theory dual \cite{AM1}, this implies that 
no superpotential is generated on the $F$-theory side.
As F-theory 3-branes correspond to Heterotic instantons, one again expects no
superpotential to
be generated in  Heterotic theory on the self-mirror $CY_3(11,11)$ based on the 
${\cal N}=2$ type IIA/Heterotic
dual of Ferrara et al 
where the same self-mirror Calabi-Yau figured on the type IIA side and the self-mirror nature was argued to show
that there are no world-sheet or space-time instanton corrections to the 
classical moduli space.

If the abovementioned triality is correct, then one 
must be able to show that there is no superpotential generated
on type $IIA$ side on the freely-acting antiholomorphic involution of $CY_3(3,243)$. 

On the mirror type $IIB$ side, the  $W$ is generated from domain-wall (
$\equiv D5$-branes wrapped around supersymmetric 3-cycles $\hookrightarrow$ $CY_3$'s)
tention.
$W_{IIB}=\int_{C:\partial C=\sum_i D_i}\Omega_3,$
$D_i$'s are 2-cycles corresponding to the positions of $D5$-branes or $O5$-planes, i.e., objects carrying
$D5$ brane charge. 
From the world-sheet point of view, the $D5$ branes correspond to disc amplitudes and
$O5$-planes correspond to ${\bf RP}^2$ amplitudes.
 As there are no branes in our theory, we need to consider only ${\bf RP}^2$
amplitudes. 
Now, type $IIA$ on a freely acting involution of a Calabi-Yau with no branes or fluxes can still
generate a superpotential because it is possible that free involution on type $IIA$ side corresponds to
orientifold planes in the mirror type $IIB$ side, which can generate a superpotential. 

The same can also be studied using localization techniques in 
unoriented closed string enumerative geometry \cite{DFM}.
Consider an orientation-reversing diffeomorphism $\sigma:\Sigma\rightarrow\Sigma$, 
an antiholomorhpic involution on the Calabi-Yau $X$ $I:X\rightarrow X$ and an
equivariant map $f:\Sigma\rightarrow X$ [satisfying $f\circ\sigma=I\circ f$]),
then the quotient spaces in $\tilde{f}:
{\Sigma\over<\sigma>}\rightarrow{X\over I}$ possesses a dianalytic structure. 
In the
unoriented theory, one then has to sum over holomorphic and antiholomorphic instantons.
For connected ${\Sigma\over<\sigma>}$, the two contributions are the same; hence 
sufficient to consider only equivariant holomorphic maps.
One constructs one-dimensional torus action, $T$, on $X$ compatible with $I$ with
isolated fixed points. The action $T$ induces an action on the moduli space of 
equivariant holomorphic maps, and one then evaluates
 the localized contributions from the fixed
points, using an equivariant version of the the Atiyah-Bott formula, 
much on the lines of Kontsevich's work.
For a Calabi-Yau 3-fold, the virtual cycle ``$[{\bar{M}}_{g,0}(X,\beta)]^{virt}$"
is zero-dimensional, and 
one has to evaluate $\int_{\Xi_s^{virt}}{1\over e_T(N_{\Xi_s}^{virt})}$,
where $\Xi_s\equiv$ is the fixed locus in the moduli space of symmetric holomorphic maps,
and one sees that one gets a match with similar calculations based on large $N$ dualities
and mirror symmetry 

For $\int d^2\theta W_{LG}$ to be invariant under $\Omega.\omega$,
given that the measure is reflected under $\Omega$, 
$\omega:W_{LG}\rightarrow -W_{LG}.$

\underline{away from the orbifold singularities}:
Promoting the action of $\omega$ to the one on the chiral superfields: 
\begin{equation}
\label{eq:ah1}
\omega:({\cal X}_1,{\cal X}_2,{\cal X}_4,{\cal X}_5,{\cal X}_6,{\cal X}_0)
\rightarrow({\bar {\cal X}}_2,-{\bar {\cal X}}_1,{\bar {\cal X}}_4,{\bar {\cal X}}_5,
{\bar {\cal X}}_6,{\bar {\cal X}}_0),
\end{equation}
and using $Re(Y_i)=|{\cal X}_i|^2$, one gets the following action of $\omega$ on the twisted chiral superfields $Y_i$'s:
\begin{equation}
\label{eq:ahmirror}
\omega:Y_1\rightarrow Y_2+i\pi,\ Y_2\rightarrow Y_1+i\pi;\
Y_{0,4,5,6}\rightarrow Y_{0,4,5,6}+i\pi.
\end{equation}
The action of $\omega$ on $Y_{4,5,6,0}$ implies that $\omega$ acts without fixed points even on the twisted
chiral superfields, further implying that there are no orientifold fixed planes, and {\it hence no
superpotential is generated on the type $IIA$ side away from the orbifold
singularities}.

\underline{after singularity resolution}:
Writing $W=\prod_{i=0}^7a_{e_{0,...,7}}{\cal X}_i^{e_i}$ with the requirements that
$\vec{l}^{(a)}\cdot\vec{e}=0$ for $a=1,2,3$ and $e_i\leq1$\cite{VafaLG}\footnote{We thank
A.Klemm for bringing \cite{VafaLG} to our attention.}, one sees that
${\cal X}_0\prod_{i=1}^7{\cal X}_i$ is an allowed term in the superpotential.
A valid antiholomorphic involution this time can be:
\begin{equation}
\label{eq:ah2}
\omega:({\cal X}_0, {\cal X}_1, {\cal X}_2, {\cal X}_3, {\cal X}_4, {\cal X}_5, {\cal X}_6, 
{\cal X}_7)\rightarrow 
({\bar{\cal X}_0}, {\bar{\cal X}_2}, -{\bar{\cal X}_1}, -{\bar{\cal X}_3}, {\bar{\cal X}_4},
{\bar{\cal X}_5}, {\bar{\cal X}_6}, {\bar{\cal X}_7}).
\end{equation}
This on the mirror LG side again implies that one will have free actions w.r.t. 
$Y_{0,3,4,5,6,7}$ 
implying there can be no orientifold planes and {\it no superpotential (is likely to be) 
generated even after singularity resolution}.

The $M$-theory uplift of the type $IIA$ side of the ${\cal N}=1$ Heterotic/type $IIA$
dual pair of \cite{VW}, as obtained in \cite{AM1}, involves the `barely $G_2$-manifold'
${CY_3(3,243)\times S^1\over{\bf Z}_2}$. In this section we consider the $D=11$
supergravity limit of $M$-theory and construct the ${\cal N}=1, D=4$ supergravity
action, and evaluate the K\"{a}hler potential for the same.

The effect of the ${\bf Z}_2$ involution that reflects the $S^1$, $H^{1,1}(CY_3)$
and takes $H^{p,q}(CY_3)$ to $H^{q,p}(CY_3)$ for $p+q=3$, where the $CY_3$ is the one 
that figures in ${CY_3\times S^1\over{\bf Z}_2}$,
at the level of $D=11$ supergravity can be obtained by first compactifying the same
on an $S^1$, then on $CY_3$ (following \cite{BCF}) 
and eventually modding out the action by the abovementioned
${\bf Z}_2$ action.

The $D=11$ supergravity action of Cremmer et al is:
\begin{equation}
\label{eq:L11}
{\cal L}_{11}=-{1\over2}e_{11}R_{11}-{1\over48}(G_{MNPQ})^2+{\sqrt{2}\over(12)^4}
\epsilon^{M_0...M_{10}}G_{M_0...M_3}G_{M_4...M_7}C^{11}_{M_8M_9M_{10}},
\end{equation}
which after dimensional reduction on an $S^1$, gives:
\begin{eqnarray}
\label{eq:L10}
& & {\cal L}_{10}=-{1\over2}e_{10}R_{10}-{1\over8}e_{10}\phi^{{9\over4}}F_{mn}^2-{9\over16}
e_{10}(\partial_m ln\phi)^2-{1\over48}e_{10}\phi^{{3\over4}}(F_{mnpq}+6F_{[mn}B_{pq]})^2
\nonumber\\
& & -{1\over12}e_{10}\phi^{-{3\over2}}H_{mnp}^2+{\sqrt{2}\over(48)^2}\epsilon^{m_0...m_9}
(F_{m_0...m_3}+6F_{m_0m_1}B_{m_2m_3})F_{m_4...m_7}B_{m_8m_9},
\end{eqnarray}
where 
\begin{eqnarray}
\label{eq:dieufs}
& & G_{MNPQ}=\partial_{[M}C_{NPQ]}\nonumber\\
& & F_{mnpq}=4\partial_{[m}C_{npq]};\ B_{mn}=C_{mn10};\ H_{mnp}=3\partial_{[m}B_{np]};\
F_{mn}=2\partial_{[m}A_{n]};\nonumber\\ 
& & C_{mnp}=A_{mnp}+3A_{[m}B_{np]},
\end{eqnarray}
and
\begin{eqnarray}
\label{eq:dieufs11to10}
& & e_{11}\ ^A_M=\left(\begin{array}{cc}
e_{10}\ ^a_m & \phi A_M \\
0 & \phi\\ 
\end{array}\right)\nonumber\\
& & A,M=0,...,10;\ a,m=0,...,9.
\end{eqnarray}
After compactifying on a $CY_3$, one gets the following Lagrangian density:
\begin{equation}
\label{eq:S1.CY31}
{\cal L}_4={\cal L}_4^{{\rm grav}+H^0}+{\cal L}_4^{H^2}+{\cal L}_4^{H^3},
\end{equation}
e.g., after a suitable Weyl scaling:
\begin{displaymath}
\label{eq:gravredieuf}
{\cal L}_4^{\rm grav}=e\biggl[-{R\over2}-{G_{AB}\over2}\partial_\mu v^A\partial^\mu v^B
-{1\over4}{\partial_\mu({\cal V}\phi^{-3})^2\over({\cal V}\phi^{-3})^2}
+G_{\alpha{\bar\beta}}\partial_\mu Z^\alpha\partial^\mu{\bar Z}^\beta\biggr],
\nonumber
\end{displaymath}
where $A,B=1,...,h^{1,1}(CY_3)$, $\alpha,\beta=1,...,h^{2,1}(CY_3)$, 
\begin{eqnarray}
\label{eq:dieufs2}
& & G_{AB}={i\int e^A\wedge *e^B\over2{\cal V}},
{\cal V}={1\over3!}\int J\wedge J\wedge J,
G_{\alpha\bar{\beta}}=-{i\int b_\alpha\wedge{\bar b}_\beta\over{\cal V}},
\nonumber\\
& & 
\bar{b}_{\alpha\ ij}={i\over||\Omega||^2}\Omega_i^{\bar{l}\bar{k}}
\bar{\Phi}_{\alpha\ \bar{l}\bar{k}j},
\end{eqnarray}
$\Phi$ being a (2,1) form, and the $h^{1,1}$ moduli 
$M^A=\sqrt{2}v^A\phi^{-{3\over4}}$,
entering in
the variation of the metric with mixed indices and the 
$h^{2,1}$ moduli $Z_\alpha$
entering in the variation of the metric with same indices. 
one gets:
Under the freely-acting antiholomorphic involution, the $h^{1,1}$-moduli $M^A$/$v^A$ 
get projected out, $G_{AB}$ is even, and $A_\mu^A$ gets projected out.
Thus, one gets:
\begin{equation}
\label{eq:LgravZ2}
{\cal L}_4^{\rm grav}/{\bf Z}_2=e\biggl[
-{R\over2} -{1\over4}{(\partial_\mu({\cal V}\phi^{-3}))^2\over({\cal V}\phi^{-3})^2}
+G_{\alpha{\bar\beta}}\partial_\mu Z^\alpha\partial^\mu{\bar Z}^\beta\biggr].
\end{equation}
Defining 
$S\equiv\tilde{\phi}+iD-{1\over4}(\Psi+{\bar\Psi})R^{-1}(\psi+{\bar\Psi})$,
$\tilde{\phi}\equiv\sqrt{2}{\cal V}(v)\phi^{-3}$,
$\Psi_{I(\equiv0,1,...,h^{2,1})}$ 
appearing in the expansion of the real 3-form
$A_{mnp}$ in a canonical basis of $H^3$, $D$ being a Lagrange multiplier,
and
\begin{eqnarray}
\label{eq:RcalNdieuf}
& & R_{IJ}\equiv Re[{\cal N}_{IJ}],\ {\cal N}_{IJ}\equiv {1\over4}{\bar F}_{IJ}
-{(NZ)_I(NZ)_J\over(ZNZ)},\
(R^{-1})^{IJ}=2\biggl(N^{-1}({\bf 1}-{\bar K}{\bar Z} - KZ)\biggr)^{IJ},
\end{eqnarray}
where $Z^I$ and $i F_I$ are the period integrals, $N_{IJ}={1\over4}(F_{IJ}+{\bar F}_{IJ})$,
 and $K_I\equiv{\int\Omega_I\wedge{\bar\Omega}\over
\int\Omega\wedge{\bar\Omega}}$, with $\Omega_I\equiv{\partial\Omega\over\partial Z^I}$.
Here, it is assumed that the holomorphic 3-form $\Omega$ is expaneded in a canonical
cohomology basis $(\alpha_I,\beta^I)$ satisfying
\begin{equation}
\label{eq:canbasdieufs}
\int_{A^J}\alpha_I=\int\alpha_I\wedge
\beta^J=-\int_{B_I}\beta^J=-\int\beta^J\wedge\alpha_I=\delta^J_I,
\end{equation}
with $(A^I,B_I)$ being
the dual homology basis. The period integrals are then defined to be:
$Z^I=\int_{A^I}\Omega$ and $iF_I=\int_{B_I}\Omega$. Hence,
\begin{equation}
\label{eq:omegaexpdieuf}
\Omega=Z^I\alpha_I + i F_I\beta^I.
\end{equation}

For the ${\cal N}=1$ case, we work in the large volume limit of the Calabi-Yau. In this
limit, one gets:
\begin{equation}
\label{eq:actionN=1}
{\cal L}^{{\rm grav}+H^0+H^2+H^3}/{\bf Z}_2=
e\biggl[-{R\over2}-G_{AB}\partial_\mu a^A\partial^\mu a^B+G_{\alpha\bar{\beta}}\partial_\mu 
Z^\alpha\partial^\mu Z^{\bar{\beta}}-{1\over2}{(\partial_\mu\tilde{\phi})^2\over\tilde{\phi}^2}
\biggr].
\end{equation}
Hence, one gets for the ${\cal N}=1$ K\"{a}hler potential $K_{{\cal N}=1}$:
\begin{equation}
\label{eq:KahN=1}
K_{{\cal N}=1}=K[a^A,Z^\alpha]+{1\over2}ln[\tilde{\phi}].
\end{equation}
At the ${\cal N}=2$ level, if there were decoupling between the fields of the $H^3$-sector
from the other fields of the other sectors, the K\"{a}hler potential would be consisting of 
$ln\biggl[S+{\bar S}+{1\over2} (\Psi+{\bar\Psi})R^{-1}(\Psi+{\bar\psi})\biggr],$\footnote{This
however assumes that ${\partial{\bar{\cal N}}_{IJ}\over\partial z^K}=0\leftrightarrow
{1\over4}F_{IJK}-{({1\over4}F_{IRK}{\bar z}^R(N{\bar Z})_J+{1\over4}(N{\bar Z}_I
F_{JLK}{\bar Z}^L)\over({\bar Z}N{\bar Z})}+{(N{\bar Z})_I(N{\bar Z})_J\over
(2{\bar Z}N{\bar Z})^2}{\bar Z}^P{\bar Z}^QF_{PQK}=0$} from the
$H^3$-sector. From the definitin of $S$ above, one sees that:
\begin{equation} 
\label{eq:S+cc}
S+{\bar S}+{1\over2} (\Psi+{\bar\Psi})R^{-1}(\Psi+{\bar\Psi})=2\tilde{\phi}.
\end{equation}
This partially explains the appearance of $ln[\tilde{\phi}]$ in $K_{{\cal N}=1}$.

Given the action of an antiholomorphic involution on the cohomology, it is in general
quite non-trivial to figure out the action of the involution on the period integrals
using the canonical (co)homology basis of (\ref{eq:canbasdieufs}). We now discuss a 
reasonable guess for the same. From (\ref{eq:S+cc}), one sees that the RHS is reflected
under the antiholomorphic involution discussed towards the beginning of this section.
We now conjecture that on the LHS, this would imply that 
\begin{eqnarray}
\label{eq:conj1}
& & S\rightarrow-S,\
(\Psi+{\bar\Psi})^2\rightarrow(\Psi+{\bar\Psi})^2,
 R_{IJ}\rightarrow-R_{IJ}.
\end{eqnarray}
We further conjecture that $R_{IJ}\rightarrow-R_{IJ}$ is realized by
\begin{eqnarray}
\label{eq:conj2}
& & Z^I\rightarrow-{\bar Z}^I,\ iF_I\rightarrow-i{\bar F}_I\
\alpha_I\rightarrow-\alpha_I,\ \beta^I\rightarrow\beta^I.
\end{eqnarray}
One should note that given that the antiholomorphic involution is orientation reversing,
the intersection form $\int\alpha_I\wedge\beta^J$ is also reflected.
This fact, e.g., can be explicitly
seen in the real basis of $H^3(T^6,{\bf Z})$\cite{KST,AM3}:
Further, the conjecture at the level of action on the cohomology, can
also be checked for  the mirror to the quintic,
for which $h^{2,1}=1$.

\section{Connection with MQCD}
In this 
section,  we make some observations and
speculative remarks regarding the possible relationship between each of the two parts of
the paper above and Witten's MQCD.
When uplifting configurations involving $NS5$ and $D4$ branes to M-theory
that displayed the interesting properties of chiral symmetry breaking,
confinement, etc, Witten in \cite{W} showed that the world volume of the $M5$-brane
is topologically $R^{1,3}\times\Sigma$, where $\Sigma$ is a Riemann
surface (embedded in a Calabi-Yau 3-fold for $N=1$ MQCD).
Now, 
the compact $CY_3(3,243)$ has been central in the second part of this paper, 
both
at the $N=2$ and at the $N=1$ levels. The mirror to the same, $CY_3(243,3)$
as shown in \cite{Lerchetal}, 
can be expressed as a $K3$-fibration over ${\bf CP}^1$ and written as
the following degree-24 hypersurface in ${\bf WCP}^4[1,1,2,8,12]$:
\begin{equation}
\label{eq:mirrorhyps}
W(a,b,c)\equiv {1\over 24}(\zeta+{b\over\zeta}+2)x_0^{12}
+{1\over12}x_3^12+{1\over3}x_4^3+{1\over2}x_5^3
+{1\over6\sqrt{c}}(x_0x_3)^6+({a\over\sqrt{c}})^{{1\over6}}x_0x_3x_4x_5=0,
\end{equation}
where $a\equiv-{\psi_0^6\over\psi_1},\ b\equiv\psi_2^{-2},\ c\equiv-{\psi_2\over\psi_1^2},\ \psi_{0,1,2}\equiv$ three complex structure deformation parameters
entering $W(x_0,x_1,x_2,x_3,x_4,x_5;\psi_0,\psi_2,\psi_2)$. Further,
in the $x_2\neq0$ coordinate patch, ${x_1\over x_2}\equiv {\zeta^{{1\over12}}
\over b^{{1\over24}}}$ and $x_1^2\equiv x_0\zeta^{{1\over12}}$, $\zeta$ being
${\bf CP}^1$ coordinate.
By choosing a particular $\alpha^\prime$-scaling for the three
complex structure deformation parameters:
\begin{equation}
\label{eq:sclgs}
a\sim\epsilon,\ b\sim\epsilon^2, 1-c\sim\epsilon, \epsilon\equiv(\alpha^\prime)^{{3\over2}}\rightarrow0,
\end{equation}
to get the $SU(3)$ Seiberg-Witten regime,
, the corresponding hypersurface
can be rewritten 
\begin{eqnarray}
\label{eq:Riemann1}
& & (\alpha^\prime)^{{3\over2}}
(z + {\Lambda^6\over z}+2P_{A_2}(x,u,v)+y^2+w^2)+O(\epsilon^2)\nonumber\\
& & 
=\epsilon^{{3\over2}}(\prod_{i=1}^3(x-a_i(z))+y^2+w^2) + O(\epsilon^2),
\end{eqnarray}
 where one sees the Riemann surface \cite{Lerchetal}
\begin{equation}
\label{eq:Riemann2}
\Sigma:\prod_{i=1}^3(x-a_i(z)).
\end{equation}
In addition, chiral symmetry breaking in the model results in the formation of domain
wall separating different vacua,
 whose world-volume is topologically given by $R^{1,2}(X^{0,1,2}\times S(x^{3,4,5})$,
where $S$ is a supersymmetric 3-cycle embedded in a $G_2$-manifold that is
topologically ${\bf R}(x^3)\times({\bf R}^5(x^{4,5,6,7,8}\times S^1(x^{10})$
Complexifying the coordinates, $v=x^4+ix^5, w=x^7+ix^8, s=x^6+ix^8,
t=e^{-s}$, the boundary condition on $S$ is that as $x^3\rightarrow-\infty$,
$S\rightarrow{\bf R}\times\Sigma$ and as $x^3\rightarrow\infty$,
$S\rightarrow{\bf R}\times\Sigma^\prime$, where $\Sigma:w=\zeta v^{-1},\
t=v^n$ and $\Sigma^\prime:w=e^{{2\pi i\over n}}\zeta v^{-1}, t=v^n$.
Following the notations of \cite{Vol1}, the calibration for $G_2$ manifolds
can be written as: $\Phi=e^{123}+e^{136}+e^{145}+e^{235}-e^{246}+e^{347}
+e^{567}$ (slightly different from (\ref{eq:caldieuf}); $e^{ijk}\equiv 
e^i\wedge e^j\wedge e^k$), and then the
supersymmetric 3-cycle embedded in the $G_2$-manifold will be given as:
$w=w(x^3,v,{\bar v}), s=s(x^3,v,{\bar v})$. Then with the choice of
vielbeins as:$e^1=dx^{10}, e^2=dx^5, e^3=dx^3, e^4=dx^7, e^5=dx^4, e^6=dx^6, 
e^7=dx^8$ and $x^6=A, x^7=C, x^8=D, x^{10}=B$, the condition for
supersymmetric cycle: $\Phi|S=\sqrt{g}dx^3\wedge dx^4\wedge dx^5$,
after further relabeling $x^{3,4,5}$ as $y^{3,1,2}$ and after assuming:
$\partial_1A=\partial_2B,\ \partial_2A=-\partial_1B(\equiv$ Cauchy-Riemann
condition), translates to give:
\begin{eqnarray}
\label{eq:susyG2a}
& & [\partial_1A\partial_3A-\partial_2\partial_3B+\partial_1C\partial_3C
-\partial_2C\partial_3D]^2+[\partial_2A\partial_3A+\partial_1A\partial_3B
+\partial_2C\partial_3C+\partial_1C\partial_3D]^2\nonumber\\
& & = [1+(\partial_1A)^2+(\partial_2A)^2+(\partial_1C)^2+(\partial_2C)^2]
[(\partial_3A)^2+(\partial_3B)^2+(\partial_3C)^2+(\partial_3D)^2].
\end{eqnarray}
The ansatz to solve (\ref{eq:susyG2a}) taken in \cite{Vol1} was:
\begin{eqnarray}
\label{eq:susyG2b}
& & A(y_1,y_2,y_3)=-ln(y_1^2+y_2^2)+\sum_{m=1}^\infty{2\over cosh y_3}^{2m} 
a_{2m},\nonumber\\
& & B(y_1,y_2,y_3)=
-2tan^{-1}({y_2\over y_1})+\sum_{m=1}^\infty{2\over cosh y_3}^{2m}b_{2m},
\nonumber\\
& & C(y_1,y_2,y_3)=(tanh y_3)({-y_1\zeta\over y_1^2+y_2^2}),\nonumber\\
& & D(y_1,y_2,y_3)=(tanh y_3)({y_2\zeta\over y_1^2+y_2^2}).
\end{eqnarray}
However, no explicit forms of $a_{2m}$ and $b_{2m}$ were given that would
solve (\ref{eq:susyG2a}). In \cite{Vol2}, however, choosing:
$e^1=dx^{10},\ e^2=dx^5,\ e^3=dx^3,\ e^4=dx^7,\ e^5=dx^6,\ dx^8$, and
for the $SU(2)$ embedding of the supersymmetric 3-cycle in the $G_2$-manifold,
the 
ansatz taken is:
\begin{eqnarray}
\label{eq:SU2ans}
& & v=
\biggl[e^{{y_1\over2}}+\sum_{m=1}^\infty\biggl({1\over2cosh y_3}\biggr)^{2m}
f_{2m}(y_1)\biggr]e^{iy_2},\nonumber\\
& & w=-\zeta tanh y_3\biggl[e^{-{y_1\over2}}+\sum_{m=1}^\infty
\biggl({1\over2cosh y_3}\biggr)^{2m}
g_{2m}(y_1)\biggr]e^{-iy_1},\nonumber\\
& & s=-y_1-\sum_{m=1}^\infty\biggl({1\over2cosh y_3}\biggr)^{2m}
h_{2m}(y_1)-2iy_2,
\end{eqnarray}
where for the $SU(2)$ group, $f_{2m},g_{2m},h_{2m}$ can be complex, but
are taken to be real in \cite{Vol2}. The condition for getting a supersymmetric
3-cycle implemented by ensuring that the pull-back of the calibration $\Phi$
to the world volume of the 3-cycle is identical to the volume form on the
3-cycle, gives recursion relations between the coefficients $f_{2m}$ and
$g_{2m}$, by setting $h_{2m}=0$, e.g. for $m=1$, as shown in \cite{Vol2},
\begin{eqnarray}
\label{eq:recursionm=0}
& & -\zeta e^{-{y_1\over2}}\partial_1f_2+(2e^{y_1\over2}
+{\zeta\over2} e^{-{y_1\over2}}f_2-\zeta e^{{y_1\over2}}\partial_1g_2-(2\zeta^2
e^{-y_1\over2}+{\zeta\over2}e^{y_1\over2})g_2=-4\zeta^2 e^{{-y_1\over2}},
\nonumber\\
& & -(\zeta^2e^{-y_1}+4)f_2+2\zeta\partial_1g_2-(\zeta^2-\zeta)g_2=-2\zeta^2
e^{-y_1\over2}.
\end{eqnarray}
One can subsitute for $f_2$ from the second equation and get a second
order differential equation for $g_2$. However, it is shown that in the limit
$\zeta\rightarrow0$, one can consistently set $f_{2m}=h_{2m}=0,m\geq1$.
Further, surprisingly, as perhaps missed to be noticed in \cite{Vol2}, 
one also gets the following differential equation for all $g_{2m}$'s, $m\geq1$:
\begin{equation}
\label{eq:zeta0}
2\partial_1g_{2m}+g_{2m}={\cal O}(\zeta)\rightarrow0,
\end{equation}
implying
\begin{equation}
\label{eq:zeta0sol}
g_{2m}=e^{-y_1\over2},\ m\geq1.
\end{equation}
Hence, (\ref{eq:SU2ans}) becomes:
\begin{eqnarray}
\label{eq:SU2ans1}
& & v(y_1,y_3)=e^{y_1\over2+iy_2},\nonumber\\
& & w(y_1,y_3)=-\zeta {tanh(y_3) e^{-y_1\over2}\over1-({
sech(y_3)\over2})^2} e^{-iy_2},
\nonumber\\
& & s(y_1,y_3)=-y_1-2iy_2.
\end{eqnarray}
One thus gets a convergent solution, unlike the case for finite $\zeta$ as
pointed out in \cite{Sonnenetal}.

Also, Seiberg-Witten equations for $N=2$ MQCD are extensively
used in the analysis. 
Now, by compactifying $E_8\times E_8$ Heterotic string theory on a 
$T^2$ at the complex structure modulus equal to the K\"{a}hler modulus equal to $i$,
there is an enhanced $SU(2)\times SU(2)$ gauge symmetry and by embedding of $SU(2)$ gauge bundles
in the two $E_8$'s and one of the two $SU(2)$'s and subsequent Higgsing away of the 
resultant $E_7\times E_7$, one gets  the (Heterotic) string analog of $N=2$ Seiberg-Witten 
theory \cite{KV}.

Hence, we see a connection between the first two parts of this paper 
and MQCD via the existence of supersymmetric 3-cycles embedded 
in $G_2$-manifolds
(around which $M2$ branes wrapped to produce membrane instantons), and
the Riemann surface that is common to both the world-volume of the $M5$ brane
used to reproduce the type $IIA$ brane configurations, and the compact
Calabi-Yau $CY_3(3,243)$ that appeared ubiquitously in Section {\bf 3}.

\section{Conclusion and Discussions}
In this paper, we have evaluated in a closed form, the exact expression for the
nonperturbative contribution to the superpotential from a single $M2$-brane 
wrapping an isolated supersymmetric 3-cycle of a $G_2$-holonomy manifold. 
The comparison with Harvey and Moore's result, is 
suggestive but not exact. A heat-kernel asymptotics analysis
for a non-compact smooth $G_2$-holonomy manifold that is metrically
${\bf R}^4\times T^3$, in the adiabatic approximation, showed that
the UV-divergent terms of one of the bosonic and the fermionic determinants
are proportional to each other, to the order we calculate, 
indicative of cancellation between the same, as expected because
the $M2$ brane action of Bergshoeff, Sezgin and Townsend
is supersymmetric. {\it Unlike the result of \cite{harvmoore},
the expression obtained for the superpotential above in terms of
fermionic and bosonic determinants, in addition to a holomorphic phase 
factor, is valid even for non-rigid supersymmetric 3-cycles as the
one considered in (\ref{eq:G2metric}) above.} 
Following Gubser et al in \cite{C}, it is 
tempting to conjecture that the superpotential term corresponding to
multiple wrappings of the $M2$-brane around the supersymmetric 3-cycle, should be give by:
\begin{equation}
\label{eq:mults}
\Delta W=\sum_n\sqrt{{det{\cal O}_3\over\det{\cal O}_1det{\cal O}_2}}{e^{n\int_\Sigma[iC-
{1\over l_{11}^3} vol(h)]}\over n^2}.
\end{equation}
In terms of relating the result obtained in (\ref{eq:Wfinal}) to that of
the 1-loop Schwinger computation of $M$ theory and the large $N$-limit of the
partition function evaluated in \cite{GopVafaMth1}\footnote{This
logic was suggested to us by R.Gopakumar.}, one notes that
the 1-loop Schwinger computation also has as its starting point, an 
infinite dimensional
bosonic determinant of the type $det\biggl((i\partial-e A)^2
-Z^2\biggr)$, $A$ being the gauge field
corresponding to an external self-dual field strength, and $Z$ denoting the 
central charge. The large $N$-limit of the partition function of Chern Simons theory
on an $S^3$, as first given by Periwal in \cite{Periwal}, involves the product of infinite
number of $sin$'s, that can be treated as the eigenvalues of an infinite determinant. 
This is indicative 
of a possible connection between the membrane instanton contribution
to the superpotential, the 1-loop Schwinger computation and the large $N$ limit
of the Chern-Simons theory on an $S^3$. Also, there were interesting
similarities and differences between the membrane instanton result of 
Section {\bf 2} and Witten's heterotic world-sheet instanton result in
terms of the forms of expressions and the boson-fermion determinant 
cancellation in both.

We also related the ${\cal N}=1$ Heterotic theory on a self-mirror 
$CY_3$ to the nonperturbative formulations of type IIA and IIB, namely 
M and F theories. While on the M-theory side, the suitable manifold turned out to 
one of $SU(3)\times {\bf Z}_2$ holonomy, referred to as a `barely $G_2$ manifold',
the elliptically fibered Calabi-Yau 4-fold involves a trivial ${\bf CP}^1$-fibration
over the Enriques surface for its base, and surprisingly has a Hodge data that can
not be obtained as a free involution of (${\cal N}=2$ F-theory on) $CY_3(3,243)\times T^2$.
The precise construction of the $CY_4$ used in the F-theory dual 
and its connection with the ${\cal N}=2$ parent model of F-theory on $CY_3(3,243)\times T^2$, 
needs to be understood.  

We also 
obtained the Meijer basis of solution to the Picard-Fuchs equation for the Landau-Ginsburg
model corresponding to $CY_3(3,243)$ {\it after} the resolution of the orbifold singularities
of the degree-24 Fermat hypersurface in ${\bf WCP}^4[1,1,2,8,12]$, 
in the large {\it and} small complex structure limits, getting the $ln$-terms 
without resorting to parametric differentiations of infinite series. 
We also discussed in detail the evaluation of the monodromy matrix in the large 
complex structure limit. We also considered the action of an antiholomorphic 
involution on 
$D=11$ supergravity compactified on $CY_3(3,243)\times S^1$, and evaluated the form of the
${\cal N}=1$ K\"{a}hler potential. In the process, we also gave a conjecture on the
action of the involution on the periods of $CY_3(3,243)$, given its action on the
cohomology of the same. We verified the conjecture for $T^6$ for the periods and
cohomology basis, and for the mirror  to the quintic for the cohomology basis. Finally, we 
showed that no superpotential is generated in type IIA and hence $M$-theory
sides using mirror symmetry, {\it after} the resolution of the orbifold singularities
associated with the Fermat hypersurface whose blow up gives $CY_3(3,243)$.

The reason for considering membrane instantons involving $M2$ branes 
wrapping around supersymmetric 3-cycles embedded in $G_2$-holonomy manifold
and topics related to the compact Calabi-Yau $CY_3(3,243)$ in the same
article was because we have tried to attempt to make an indirect
connection between these two topics by relating both of them individually
to Witten's MQCD. The latter involves uplifting suitable configurations of
$NS5$ and $D4$ branes to $M$ theory involving $M5$ branes with a suitable
topology, as well as domain walls being modelled again by $M5$ branes with
a particular topology, the former case consisting of a Riemann surface
that is also present in the defining hypersurface of $CY_3(3,243)$, and
the latter corresponding to precisely a supersymmetric 3-cycle embedded
in a $G_2$-holonomy manifold. We also showed that in the limit of
vanishing of a certain complex constant that figures in the Riemann surface
when referring to the boundary conditions satisfied by the supersymmetric
3-cycle embedded in $G_2$-manifold, it was possible to get an exact 
answer for the embedding, using the ansatz of \cite{Vol2}.

\end{document}